\documentclass[fleqn,usenatbib]{mnras}

\usepackage[T1]{fontenc}
\usepackage{pdflscape}

\DeclareRobustCommand{\VAN}[3]{#2}
\let\VANthebibliography\thebibliography
\def\thebibliography{\DeclareRobustCommand{\VAN}[3]{##3}\VANthebibliography}

\usepackage{graphicx}	% Including figure files
\usepackage{amsmath}	% Advanced maths commands
\usepackage{amssymb}	% Extra maths symbols
\usepackage{arydshln}
\usepackage{multirow}

\usepackage{gensymb}

\def \aj {AJ}
\def \mnras {MNRAS}
\def \apj {ApJ}
\def \apjs {ApJS}
\def \apjl {ApJL}
\def \aap {A\&A}
\def \nat {Nature}
\def \araa {ARAA}
\def \pasp {PASP}
\def \aaps {AAPS}

\usepackage{newtxtext,newtxmath}
\title[Low-frequency view of HMXBs]{Low-frequency spectra of neutron star + OB supergiant binaries: \\ \textit{Does wind density drive persistent and flaring modes of accretion?}}

\author[J. van den Eijnden et cal.]{J. van den Eijnden$^{1,2}$\thanks{a.j.vandeneijnden@uva.nl}
L. Sidoli,$^{3}$
M. Diaz Trigo,$^{4}$
I. El Mellah,$^{5,6}$
V. Sguera,$^{7}$
N. Degenaar,$^{1}$
F. F\"urst,$^{8}$
\newauthor 
V. Grinberg,$^{9}$
P. Kretschmar,$^{10}$
S. Mart\'inez-N\'u\~nez,$^{11}$ 
J. C. A. Miller-Jones,$^{12}$
K. Postnov,$^{13,14}$
T. D. Russell$^{15}$
\\
$^{1}$Anton Pannekoek Institute for Astronomy, Universiteit van Amsterdam, Science Park 904, 1098, XH, Amsterdam, The Netherlands\\
$^{2}$Department of Physics, University of Warwick, Coventry CV4 7AL, UK\\
$^{3}$INAF, Istituto di Astrofisica Spaziale e Fisica Cosmica, Via A. Corti 12, 20133, Milano, Italy\\
$^{4}$ESO, Karl-Schwarzschild-Strasse 2, 85748, Garching bei M\"unchen, Germany\\
$^{5}$Departamento de F\'isica, Universidad de Santiago de Chile, Av. Victor Jara 3659, Santiago, Chile \\
$^{6}$ Center for Interdisciplinary Research in Astrophysics and Space Exploration (CIRAS), USACH, Chile\\
$^{7}$ INAF, Osservatorio di Astrofisica e Scienza dello Spazio, Via P. Gobetti 101, I-40129, Bologna, Italy\\
$^{8}$ Quasar Science Resources SL for ESA, European Space Astronomy Centre (ESAC), Science Operations Departement, 28692, Villanueva de la Cañada, Madrid, Spain\\
$^{9}$ European Space Agency (ESA), European Space Research and Technology Centre (ESTEC), Keplerlaan 1, 2201 AZ, Noordwijk, The Netherlands\\
$^{10}$ European Space Agency (ESA), European Space Astronomy Centre (ESAC), Camino Bajo del Castillo s/n, 28692, Villanueva de la Cañada, Madrid, Spain\\
$^{11}$ Instituto de Física de Cantabria (CSIC-Universidad de Cantabria), 39005, Santander, Spain\\
$^{12}$ International Centre for Radio Astronomy Research, Curtin University, GPO Box U1987, Perth, WA 6845, Australia\\
$^{13}$ M.V. Lomonosov Moscow State University, Sternberg Astronomical Institute, 13, Universitetskij pr., 119234, Moscow, Russia\\
$^{14}$ Kazan Federal University, Kremlevskaya 18, 420008 Kazan, Russia\\
$^{15}$ Istituto di Astrofisica Spaziale e Fisica Cosmica, INAF, Via U. La Malfa 153, Palermo, I-90146, Italy\\
}

\date{Accepted XXX. Received YYY; in original form ZZZ}

\pubyear{2023}

\begin{document}
\label{firstpage}
\pagerange{\pageref{firstpage}--\pageref{lastpage}}
\maketitle

% Abstract of the paper
\begin{abstract}
Neutron star high-mass X-ray binaries are well-studied in wavebands between the infrared and hard X-rays. Their low-frequency millimeter and radio properties, on the other hand, remain poorly understood. We present observations of the millimeter and radio emission of binaries where a neutron star accretes from an OB supergiant. We report ALMA and NOEMA millimeter observations of twelve systems, supplemented by VLA radio observations of six of those targets. Our targets include six Supergiant X-ray Binaries (SgXBs), four Supergiant Fast X-ray Transients (SFXTs), and two intermediate systems. Nine out of twelve targets, including all SFXTs, are detected in at least one millimeter band, while in the radio, only two targets are detected. All detected targets display inverted radio/millimeter spectra, with spectral indices in the range $\alpha =0.6-0.8$ for those systems where accurate SED fits could be performed. We conclude, firstly, that the low-frequency SEDs of neutron star SFXTs and SgXBs are dominated by free-free emission from the OB supergiant's stellar wind, and that jet emission is unlikely to be observed unless the systems can be detected at sub-GHz frequencies. Secondly, we find that SFXTs are fainter at 100 GHz than prototypical SgXBs, probably due to systematically less dense winds in the former, as supported further by the differences in their fluorescence Fe K$\alpha$ lines. We furthermore compare the stellar wind constraints obtained from our millimeter observations with those from IR/optical/UV studies and bow shock detections, and present evidence for long-term stellar wind variability visible in the thermal emission.
\end{abstract}


\section{Introduction}

Massive stars ($M \gtrsim 8M_{\odot}$) play a central role in a wide range of astrophysical processes. As the rare, high end of the stellar mass scale, massive stars are key in understanding stellar formation and evolution. Given their prevalence to reside in binary or higher-order systems \citep{sana2012}, massive stars drive binary evolution towards its energetic intermediate (X-ray binaries) and end states (mergers of binary compact objects). The endpoints of massive star evolution power a range of cataclysmic transients, such as supernovae and long gamma-ray bursts that may lead to the formation of stellar-mass black holes and neutron stars, and kilo-novae in the case of binary neutron star mergers \citep{eldridge2004,belczynski2010}. 

Evolved massive stars drive powerful radiative and mechanical feedback on a range of scales, from their directly surrounding interstellar medium (ISM) to their host galaxies. Their mechanical feedback takes the form of stellar winds, that play a role in many of the earlier mentioned processes -- affecting massive star evolution up to the transients they power, the evolution of the binaries that host them, or the mass distribution of resulting neutron stars and black holes \citep{laplace2025}. These winds shape and energize the surrounding ISM, powering particle acceleration in local shocks around individual stars \citep[e.g.,][]{dopita1990,delvalle2012,delpalacio2018,Prajapati2019,meyer2020}. On scales of young stellar clusters, stellar winds can combine into a cluster wind to create collective feedback structures in the surrounding medium --  where it may power the (re-)acceleration of cosmic rays, as suggested by recent $\gamma$-ray detections of such clusters \citep[e.g.,][]{Aharonian2019,peron2024,aharonian2024} and acceleration modeling \citep{harer2023,vieu2023,menchiari2025}. Moving up further in scale, massive stars and their winds contribute to the heating and structure of their host galaxy \citep{andersson2020}. 

An observational understanding of both the global wind properties and its micro-structure are indispensable to understand this range of feedback and interaction processes. Within the wind, its density structure (where inhomogeneities are often referred to as clumping) and velocity profile affect, for instance, the accretion of the wind by a companion \citep[e.g.,][]{bozzo2016,elmellah2018,bozzo2021,ramachandran2025}. On larger scales, the global mass-loss rate $\dot{M}_{\rm w}$ and terminal velocity $v_\infty$ of the stellar wind set the power budget for the feedback on -- and particle acceleration in -- the ISM and stellar cluster surroundings. 

A wide range of methods exists to measure the global and micro wind properties \citep[see e.g.,][for reviews]{puls2008,vink2022}. The overall mass-loss rate and velocity profile, for instance, can be studied via P-Cygni profiles in the UV or via the H$\alpha$ recombination line, asymmetry in X-ray emission lines, or the distance between the star and the bow shock it drives in the ISM when moving at supersonic speeds. All but the latter of these methods are affected by the clumping of the winds, as the inferred mass-loss rates depend non-linearly on density \citep[e.g.][]{hamann1988,puls2006,puls2008,sundqvist2013}. When not taken into account, clumping may therefore lead to discrepancies between inferences from different methods \citep{fullerton2006}. Variability in the wind, combined with non-simultaneous measurements, may further exacerbate such differences \citep{haucke2018,massa2024}. The bow shock method does not suffer from such effects -- clumping has smoothed out at these parsec-scale distances from the star and the shock's radiative time scales far exceed the relevant variability time scales \citep{mohamed2012}. However, it remains limited in its application due to the required knowledge about the ISM density and temperature \citep{wilkin1996,martinez2023}, and its need for supersonic peculiar motion.

On large scales within the wind, where it has reached its terminal velocity, free-free (Bremsstrahlung) thermal emission dominates the wind's spectral energy distribution (SED) at radio and sub-millimeter wavelengths. As clumping is expected to decrease as a function of radius, this low-frequency continuum emission is expected to be least affected by small-scale density structures \citep{lamers1984,daley2016,rubiodiez2022}. For a smooth wind, it was analytically derived in independent papers that this process should lead to a power-law low-frequency spectrum with a slope close to $\alpha = 0.6$ \citep{wright1975,panagia1975}, where the flux density $S_\nu$ scales with frequency $\nu$ as $S_\nu \propto \nu^\alpha$. Any residual clumping in the outer wind is expected to affect the spectrum's flux density scaling by maximally $50$\% in the millimeter band and to a lesser extent at radio wavelengths \citep{daley2016}. 

In addition to clumping, stellar winds from early-type OB-stars can be noticeably magnetized. Indeed, the surface dipole magnetic field in about 10\% of OBA-stars may reach kG values \citep{OBstars_Bfield17,Bstar_Bfield19}. The magnetized stellar winds can significantly affect the particle acceleration in wind collision zones in gamma-ray binaries with pulsars \citep{Bykov+24a} and be responsible for triggering of SFXT outbursts  by magnetic reconnection near neutron star magnetospheres \citep{2014MNRAS.442.2325S}. 

Low-frequency observations of massive star winds have predominantly been performed before the current generation of radio and millimeter interferometers \citep[e.g.,][]{bieging1989,leitherer1991,leitherer1995,blomme2003,benaglia2007a,benaglia2007b} with typical detection limits of hundreds of $\mu$Jy to several mJy. Current instrumentation is capable of significantly deeper observations: in radio, such instruments include the upgraded Karl G. Jansky Very Large Array (VLA), the upgraded Australia Telescope Compact Array (ATCA), and Square Kilometer Array (SKA) precursors such as MeerKAT and the Australian SKA Pathfinder (ASKAP). In the (sub-)millimeter band, the current state of the art is provided by the Northern Extended Millimeter Array (NOEMA) and the Atacama Large Millimeter/submillimeter Array (ALMA). \citet{fenech2018}, \citet{andrews2019}, and \citet{gallego2021} have recently showcased the power of combining VLA and ATCA radio with ALMA millimeter observations of Westerlund 1 and the Arches cluster to study their massive stellar population. However, as studies at the enhanced sensitivity of these instruments remain rare, new applications of the low-frequency approach to understand global stellar wind characteristics is warranted \citep{vink2022}.

\subsection{Probing massive star winds with millimeter, radio, and X-ray observations of accreting neutron star binaries}

When massive stars reside in a binary system with a compact object companion, the interaction between the stellar wind and the compact object can provide an orthogonal view on the wind's global and small-scale structure. The nature of the compact object in these binaries is most often a neutron star. Young, energetic neutron stars that still launch a powerful pulsar wind are thought to power a subset of so-called $\gamma$-ray binaries, whose broad-band SED is dominated by the non-thermal shock between stellar and pulsar wind \citep[e.g.][]{dubus2013}. More commonly, the closely-orbiting neutron star spins slowly ($\gtrsim1$ second period) and does not launch a pulsar wind \citep{reig2011,vandeneijnden_RNAAS2024}. A a result, the interaction with the massive star instead takes the form of accretion, as the neutron star gravitationally captures the stellar wind, in systems named High-Mass X-ray Binaries (HMXBs). This accretion process results in X-ray emission, whose luminosity can provide an independent constraint on the stellar wind properties, especially if information about the orbit is known. Alternatively, varying X-ray absorption along the binary can reveal the overall density structure, as well as clumpiness, of the stellar wind \citep{grinberg2015,elmellah2020,diez2022}. Low-frequency observations of the thermal stellar wind emission trace the wind far beyond typical binary separation \citep{gudel2002}; stellar wind accretion by the neutron star, on the other hand, probes the wind's inner regions. For eccentric orbits, furthermore, different wind regions are sampled as the neutron star moves towards and away from the star. All combined, the radio, millimeter, and X-ray properties of the HMXB along its orbit provide complementary views of the wind's global and micro-structure. 

While the presence of a neutron star companion provides an orthogonal probe of the stellar wind, it also complicates the system in two manners: the launch of relativistic jets affecting the low-frequency methods and the introduction of intrinsically complex accretion behavior affecting the X-ray methods. The effect of jets is seen most clearly in the black hole -- massive star binary Cyg X-1, where the jet dominates at radio and millimeter frequencies over the stellar wind's large-scale emission and strongly dynamically interacts with the stellar wind \citep{fender2000_cygx1,zdziarski2012,millerjones2021}. The radio jets of slowly-spinning neutron stars are observed to be significantly weaker than in Cyg X-1 \citep{vandeneijnden2018,vandeneijnden2021}, but their interaction with stellar winds may brighten them \citep{vandeneijnden2024_lsv4417}. Therefore, the potential contribution of such jets to the low-frequency emission of neutron star HMXBs should be taken into account. In a similar fashion, other non-thermal processes -- the interaction of clumps with large-scale wind structures induced by the orbiting neutron star -- may contribute at low frequencies, although such interactions have not been observed directly in HMXBs yet \citep{vandeneijnden2021}.

The complex accretion behavior complicates any inferences about the stellar wind made from X-ray observations. In this context, the different modes of accretion seen for neutron stars orbiting OB supergiants are particularly noteworthy. HMXBs hosting OB supergiants show an apparent dichotomy in X-ray behavior: the Supergiant X-ray Binaries (SgXBs) that accrete persistently on one hand, and the Supergiant Fast X-ray Transients (SFXTs) that accrete in a transient, flaring manner on the other. The origin of this difference in X-ray behavior is poorly understood, but may be driven by the properties of the neutron star \citep[e.g.,][]{shakura2012,shakura2014}, donor star and its stellar wind \citep{negueruela2019,gimenez2016,hainich2020}, binary orbit, or complex combinations of the above \citep[see e.g.,][for recent reviews]{Sidoli2017,Kretschmar2019}. While the difference between the accretion in SFXTs and SgXBs complicates using the X-ray observations as a probe of the wind, the argument can be turned around: better understanding the winds, for instance via low-frequency observations, can also test the role of the wind properties in driving accretion modes.

Published low-frequency constraints on neutron star HMXBs, particularly SFXTs, remain sparse. \citet{vandeneijnden2021} presented ATCA radio observations of eleven SgXBs, detecting emission from five systems. These radio-band-only observations, however, did not unambiguously reveal the origin of the emission beyond doubt: stellar winds only, accretion-driven jets, other non-thermal processes, their interaction, or their superposition, all remained consistent with the observations. Millimeter observations, where the stellar wind is more likely to dominate due to its strongly inverted spectrum, are therefore crucial. A pilot NOEMA millimeter study of one SgXB and one SFXT was published by \citet{vandeneijnden2023mm}, reporting the SFXT as the first millimeter-detected neutron star accreting from a massive star. With just one source per neutron star HMXB sub-class, however, this pilot study could not yet reveal whether systematic low-frequency differences between SFXTs and SgXBs may be linked to their different accretion modes.

\subsection{Constructing a first sample of neutron star HMXBs observed at millimeter frequencies}

Here, we present the first extended sample of neutron stars accreting from massive stars, observed at millimeter and radio wavelengths. Our campaign performed millimeter observations with ALMA and NOEMA, supplemented by a supporting VLA radio campaign and X-ray observations using \textit{Swift}, \textit{MAXI}, and \textit{INTEGRAL}. In this first paper, we focus on an overview of the target sample and the low-frequency observations, presenting the detection of nine out of twelve targets at millimeter wavelengths. We then specifically discuss the dominant origin of the low-frequency emission, comparing stellar wind and relativistic jet scenarios. We subsequently discuss whether the low-frequency spectra of neutron star HMXBs contain evidence that stellar wind properties drive the difference between SFXTs and SgXBs. Finally, we briefly compare the low-frequency wind inferences with bow shock inferences for two targets, and discuss wind variability in the context of our observations. In a follow-up paper, we will focus on more detailed modelling of the wind and the accretion process in our targets, combining the low-frequency data presented in this paper with X-ray constraints and updated modelling of the stellar wind's thermal free-free emission. 

\section{Targets, observations, and data analysis}
\label{sec:data}

\setlength{\tabcolsep}{6pt} % Default value: 6pt
\begin{table*}
\caption{The target sample, listed target name, type, observatory per band, distance, and donor star type, the latter two with references. *Already published in \citet{vandeneijnden2023mm}. **Range represents the range in best-fit measurements reported by the listed references. ***While not a SgXB according to its donor classification, we will group it amongst the SgXB class due to its persistent X-ray properties akin to the other SgXBs.}
\label{tab:targts}
\begin{tabular}{llllllll}
\hline
Name & Type & MM & Radio & Distance [kpc] & Distance refs & Donor type & Donor type refs \\
\hline
4U 1700-37 & SgXB & ALMA & VLA &  $1.58\pm0.07$ & \citet{vandermeij2021} & O6.5 Iaf+ & \citet{walborn1990} \\

Vela X-1 & SgXB & ALMA & VLA &  $1.99^{+0.13}_{-0.11}$ & \citet{kretschmar2021} & B0.2 Ia & Same as distance ref\\ \hdashline

4U 1907+09 & SgXB & ALMA & VLA & $3.6\pm0.7$ & Gaia DR3 & O8/O9 Ia & \citet{vankerkwijk1989} \\
& & NOEMA & & & & & \\ \hdashline

X1908+075 & SgXB & NOEMA* & -- & $4.58\pm0.50$ & \citet{martinez2015} & B0-B3 & Same as distance ref  \\

2S 0114+650 & SgXB & NOEMA & -- & $4.5\pm0.2$ & Gaia DR3 & B1Ia & \citet{reig1996} \\

4U 2206+54 & HMXB*** & NOEMA & --  & $3.1\pm0.1$ & Gaia DR3 & O9.5V & \citet{blay2006} \\

IGR J00370+6122 & \textit{Interm.} & NOEMA & -- & $3.4\pm0.2$ & Gaia DR3 & B1 Ib & \citet{Gonzalez2014} \\

IGR J19140+0951 & \textit{Interm.} & NOEMA & --  & $4.0\pm1.7$ & Gaia DR3 & B0.5 Ia & \citet{torrejon2010} \\

IGR J17544-2619 & SFXT & ALMA & VLA & $2.52\pm0.17$ & Gaia DR3 & O9I & \citet{gimenez2016} \\ \hdashline

IGR J18410-0535 & SFXT & ALMA & VLA & $3.2$--$13.9$** & \citet{Sguera2009} & B1Ib & Same as distance refs \\
& & NOEMA* & ATCA* & & \citet{nespoli2008} & B1I & \\
& &  &  & & \citet{negueruela2008} & B0.2Ibp &  \\
& &  &  & & \citet{coleiro2013} & &  \\
& & &  & & Gaia DR3 & &  \\ \hdashline

SAX J1818.6-1703 & SFXT & ALMA & VLA & $2.1 \pm 0.1$ & \citet{torrejon2010} & B0.5Iab & Same as distance ref \\

IGR J18483-0311 & SFXT & NOEMA & -- & $2.7^{+1.5}_{-0.8}$ & Gaia DR3 & B0.5-B1 & \citet{torrejon2010} \\ \hline

\end{tabular}\\
\end{table*}

\setlength{\tabcolsep}{16pt} % Default value: 6pt
\begin{table*}
\caption{The results of the NOEMA and ALMA millimeter observations. Uncertainties in flux density are given at $1\sigma$, upper limits at $3\sigma$. Observations listed at 40, 150, and 300 GHz were performed by ALMA. Observations at 100 GHz were performed by NOEMA. For the ALMA observations, we list the start time of the first target scan in the second column. For the NOEMA observations, we list the multiple dates on which observations for that target were performed. $T_{\rm obs}$ is the total observing time, including overheads, per band, in minutes. $^{*}$ALMA observations are separated by at least one orbital period. $^{**}$NOEMA observations are separated by at least one orbital period. $^{***}$The ALMA and NOEMA observation sequences both, individually, fall within a single orbital period, but the two telescope campaigns are separated by more than an orbit.}
\label{tab:mmresults}
\begin{tabular}{llllll}
\hline
Name & Date(s)/time(s) & $T_{\rm obs}$ [min] & $\nu$ [GHz] & $F_\nu$ [$\mu$Jy] & $L_\nu$ [erg/s/Hz] \\% $\alpha_{\rm MM}$ \\
\hline
\multirow{3}{*}{4U 1700-37$^{*}$} & 2024-10-05 20:57:02 & 30 & 40 & $ 1417\pm12 $ & $(4.2 \pm 0.4)\times10^{18}$ \\% & \multirow{3}{*}{$0.66 \pm 0.01$}\\
 & 2024-10-19 22:54:44 & 25 & 150 & $ 3729\pm21 $ & $(11.1 \pm 1.0)\times10^{18}$ \\% &  \\
  & 2024-10-01 21:34:03 & 38 & 300 & $ 5508\pm28 $ & $(16.5 \pm 1.5)\times10^{18}$ \\ \hdashline %  & \\ \hdashline
  
\multirow{3}{*}{Vela X-1}& 2024-10-05 11:42:35 & 29 & 40 & $ 375\pm12 $ & $(1.8 \pm 0.2)\times10^{18}$ \\% & \multirow{3}{*}{$0.78 \pm 0.01$} \\
 & 2024-10-05 12:31:45 & 24 & 150 & $ 994\pm22 $ & $(4.7 \pm 0.6)\times10^{18}$ \\% & \\
  & 2024-10-04 12:30:04 & 37 & 300 & $ 1775\pm30 $ & $(8.4 \pm 1.1)\times10^{18}$ \\ \hdashline % & \\ \hdashline
  
\multirow{4}{*}{4U 1907+09$^{**}$}& 2024-11-04 22:14:56 & 24 & 40 & $ 40\pm19 $ & $(6.2\pm3.8)\times10^{17}$ \\% & \multirow{4}{*}{$0.8 \pm 0.7 / 0.3$*} \\
& 2024-10-06 23:38:27 & 27 & 150 & $ 113\pm39 $& $(1.8\pm0.9)\times10^{18}$ \\% & \\
& 2024-10-01 22:13:27 & 42 & 300 & -- & -- \\% &  \\
& 2024-02-2/4/28, 2024-03-14 & 150 & 100 & $ 57\pm13 $ & $(8.8\pm4.0)\times10^{17}$ \\ \hdashline % & \\ \hdashline

X1908+075$^{**}$& 2023-01-25, 2023-02-07 & 88 & 100 & $<34.4$& $<8.6\times10^{17}$ \\% & \\%N/A \\

2S 0114+650& 2024-01-14, 2024-01-16 & 216 & 100 & $63 \pm 11$& $(1.5 \pm 0.3)\times10^{18}$  \\% & $-1.2 \pm 2.1$ \\

4U 2206+54& 2024-03-11/19 & 108 & 100 & $ < 36$ & $<4.1\times10^{17}$ \\% &  N/A \\

IGR J00370+6122& 2024-01-14, 2024-01-16 & 216 & 100 & $<33$& $<4.6\times10^{17}$  \\% & N/A \\

IGR J19140+0951$^{**}$& 2024-02-02/4/28, 2024-03-14 & 126 & 100 & $209 \pm 12$ & $(4.0 \pm 3.4)\times10^{18}$ \\ \hdashline% & $-1.0 \pm 0.7$ \\

\multirow{3}{*}{IGR J17544-2619$^{*}$}& 2024-10-05 23:23:15 & 38 & 40 & $ <34 $& $<2.6\times10^{17}$ \\% & \multirow{3}{*}{$0.42^{+0.04}_{-0.09}$} \\
 & 2024-10-19 23:20:18 & 68 & 150 & $108 \pm 12$& $(8.2 \pm 1.4)\times10^{17}$ \\% & \\
  & 2024-10-11 19:46:19 & 61 & 300 & $108 \pm 13$& $(8.2 \pm 1.5)\times10^{17}$ \\ \hdashline % & \\ \hdashline
  
\multirow{4}{*}{IGR J18410-0535$^{***}$}& 2024-10-05 22:03:54 & 38 & 40 & $ <38 $& $<2.1\times10^{18}$ \\% & \multirow{3}{*}{$1.2 \pm 0.2$} \\
 & 2024-10-07 00:33:24 & 65 & 150 & $40 \pm 11$& $(2.3 \pm 1.3)\times10^{18}$ \\% & \\
  & 2024-10-01 22:55:08 & 63 & 300 & $86 \pm 15$& $(4.9 \pm 2.6)\times10^{18}$ \\%  & \\
   & 2023-02-20/22/23 & 88 & 100 & $63.4 \pm 9.6$& $(3.6 \pm 1.9)\times10^{18}$ \\ \hdashline % & N/A \\ \hdashline
   
\multirow{3}{*}{SAX J1818.6-1703}& 2024-10-06 21:11:21 & 37 & 40 & $123 \pm 14$& $(6.5 \pm 0.8)\times10^{17}$ \\% & \multirow{3}{*}{$0.62^{+0.02}_{-0.04}$} \\
 & 2024-10-06 22:31:27 & 65 & 150 & $288 \pm 12$& $(1.5 \pm 0.1)\times10^{18}$ \\% & \\
  & 2024-10-02 01:21:33 & 61 & 300 & $463 \pm 17$& $(2.4 \pm 0.1)\times10^{18}$ \\ \hdashline % & \\ \hdashline
  
IGR J18483-0311$^{**}$& 2024-03-20, 2024-07-11/12 & 462 & 100 & $112 \pm 9$& $(9.7 \pm 5.8)\times10^{17}$  \\% & $-0.6 \pm 0.9$ \\
\hline
\end{tabular}\\
\end{table*}

\setlength{\tabcolsep}{10pt} % Default value: 6pt
\begin{table*}
\caption{Summary of the VLA radio observations of the six ALMA targets. The date/time column refers to the start of the observation. Each observation lasted 1 hour including overheads. The third and fourth column report the flux and phase calibrator sources (the same at each frequency). Robust refers to the robustness parameter in the \textsc{casa} \textsc{tclean} task. Errors on the flux density are quoted at $1\sigma$, upper limits are quoted at $3\sigma$. *Note that the 10-GHz observation of 4U 1700-37 suffered from gain calibration issues; this flux density is therefore not used in further analysis or interpretation.}
\label{tab:appendix_radio}
\begin{tabular}{lllllll}
\hline
Name & Date / time [UTC] & Flux calibrator. & Phase calibrator & Frequency [GHz] & Robust & Flux density [$\mu$Jy] \\ \hline
4U 1700-37 & 2024-10-15 21:01:00 & 3C286 & J0922-3959 & 6 & 1 & $264 \pm 11$\\
 & & & & 10 & 1 & \textit{207 $\pm$ 87}* \\
Vela X-1 & 2024-10-17 13:04:00  & 3C286 & J1717-3948 & 6 & 1 & $73.7 \pm 10.0$ \\
 & & & & 10 & 1 & $91.0\pm15.0$ \\
4U 1907+09 & 2024-10-16 01:46:00  & 3C48 & J1912+0518 & 6 & 1 & $<24$ \\
 & & & & 10 & 1 & $<36$ \\
IGR J17544-2619 & 2024-10-17 01:05:00  & 3C48 & J1825-0737 & 6 & 0.5 & $<30.0$ \\
 & & & & 10 & 0 & $<60.0$ \\
IGR J18410-0535 & 2024-10-17 02:05:00  & 3C48 & J1751-2524 & 6 & -- & -- \\
 & & & & 10 & 0 & $<78.0$ \\
SAX J1818.6-1703 & 2024-10-19 22:45:00  & 3C286 & J1825-1718 & 6 & 1 & $<27$ \\
 & & & & 10 & 1 & $<100.0$ \\ \hline
\end{tabular}\\
\end{table*}

\subsection{Targets}

In this work, we discuss the twelve targets that have been observed at the current state-of-the-art millimeter band sensitivity (e.g., RMS sensitivities of $10$--$40$ $\mu$Jy/bm). Our target sample is summarized in Table \ref{tab:targts}, where we list the source name, type (SgXB, SFXT, or intermediate, i.e., showing signatures of both classes), the observatories used for the millimeter, radio, and X-ray observations, and the distance and donor type with references. The NOEMA millimeter observations of two targets (the SFXT IGR J18410-0535 and SgXB X1908+075) were already published in the pilot study by \citet{vandeneijnden2023mm} and are included again here. For the former system, new millimeter observations with ALMA are presented here for the first time. The full sample spans a range in accretion behaviour and donor star type, but has primarily been constructed based on observability with NOEMA / ALMA and distance. For seven targets, we list only the \textit{Gaia} DR3 distance, providing the best constraint. For four targets, other works using (a combination of) different methods provide a better constrained distance measurement than \textit{Gaia}, which we adopt instead. 

For IGR J18410-0535, we list a range in distance estimates, as this property is poorly constrained (see Van den Eijnden et al. 2023, for the impact this has on further understanding of its millimeter emission). For 4U 2206+54, we list HMXB as binary type, as its donor is not a supergiant but a O9.5V star \citep{blay2006}. In our Figures, we group it with the SgXB as it persistently accretes from the donor star's wind. However, we will note its different nature wherever relevant. Finally, we note that 4U 1907+09 is listed as SgXB, but shows signatures of SFXTs-like X-ray dipping and has therefore been discussed as a source in-between both classes as well \citep{doroshenko2012}.

\subsection{Millimeter observations}

\subsubsection{NOEMA observations}

Six targets, including one overlapping with the ALMA targets (4U 1907+09), were newly observed with NOEMA in a single-band detection experiment at 100 GHz. The observing program (number w23bo) consisted of four tracks, with one or two targets per track; two targets were combined in a single track if their proximity allowed the use of shared calibrators to optimize overheads. Each track was observed across multiple dates to achieve the required sensitivity. The exact dates per target are listed in Table \ref{tab:mmresults}. The observations were performed in a continuum-only point-source-detection setup, where the observing band is split into a Lower Side Band (LSB; 82.5-90 GHz) and Upper Side Band (USB; 98-105.5 GHz) with centers separated by 15.488 GHz. All observations consisted of a standard scan setup, with multiple calibrators used for bandpass, amplitude, and phase calibration; the latter two were performed with one or multiple calibrators interleaving the target scans. 

We calibrated the data using the \textsc{Continuum and Line Interferometric Calibration} (\textsc{clic}) software within the \textsc{gildas} package (version Jan 2025), designed specifically for the analysis of NOEMA observations. The calibration was performed using the standard pipeline, which handles bandpass, phase, and amplitude calibration, as well as automated flagging using the pre-defined limits for detection experiments. The pipeline output was manually inspected and combined with observing log information to prepare and optimize re-runs of the pipeline, until satisfactory calibration was obtained. For observations in A configuration -- NOEMA's most extended configuration, used between 2024 February 28 and March 20 -- acceptable calibration required switching from antenna to baseline-based configuration for bandpass and amplitude. Calibration was performed separately per observing date; for the 2024 March 19 observations, the raw data was split into three parts as one antenna dropped out and came back online during the observations. Lacking a bandpass calibrator or sufficient exposure of the other calibrators, the second and third segment could not be adequately calibrated and were therefore dropped from further analysis. Similarly, issues with the bandpass calibration prevented the use of data collected on 2024 March 14. Finally, the calibrated datasets were converted to UV-tables (\textsc{.UVT} format) for further analysis, with one UV-table per target per side band. All pipeline reports, and pipeline logs in plain-text \textsc{.clic} format, are available in this paper's Data Reproduction Package. 

We performed cleaning and imaging of the observations using the \textsc{mapping} software within \textsc{gildas}. For each observation, we first performed the analysis per side band, to assess whether the target was significantly detected in each band, before combining the two UV-tables into a single band. For this analysis, we frequency-averaged the UV-tables using \textsc{uv\_cont} and cleaned across a region with size twice the full-width-half-maximum of the primary beam, down to a flux density level of 2 times the expected RMS noise (2.5 times for data including A configuration). After inspecting the resulting image and measuring the RMS noise with the \textsc{go noise} routine, we fitted a point source model in the UV plane with initial position estimated at the phase centre. While the detected targets are all visibile in their respective side-band images, we use the UV fit outcome as the formal assessment of detection: for three out of six sources, the fit results in a significant source at the phase centre, while for the two non-detected sources, a flux density below the RMS noise is found. Finally, for 4U 1907+09, this procedure is performed at a small offset from the phase centre, due to a slight pointing offset, also leading to a target detection. Finally, we merge the side-band UV-tables using \textsc{uv\_merge} for a final RMS measurement and flux density measurement for detected sources. In Table \ref{tab:mmresults}, we list the detected flux density in the combined 100-GHz band for four detected targets, and the three sigma upper limit for the two non-detected targets.

\subsubsection{ALMA observations}

An additional six targets were observed with ALMA in a multi-band observing campaign (under code 2024.1.00657.S). The targets, three SgXBs and three SFXTs, were each observed at Bands 1 (40 GHz), 4 (150 GHz), and 7 (300 GHz). The exposure times were set up to optimize sensitivity at the lower frequencies, where an inverted stellar-wind type spectrum is expected to be faintest. All 18 observations were successfully performed between 2024 Oct 01 and Nov 04 and reached the intended RMS sensitivity. 

We performed further analysis using the standard ALMA pipeline images for each target and (sub) band. Specifically, we first analysed the primary-beam-corrected continuum image for each full band, making use of self-calibrated images when available (i.e., when the target was sufficiently bright). We converted each \textsc{fits} image into the \textsc{.image} format of the Common Astronomy Software Application (CASA v.6.6.5.31) through its \textsc{importfits} tool. Those images were subsequently fitted with a 2D Gaussian point source in the image plane, fixing the Gaussian's shape to the major and minor axes and position angle of the synthesized beam. We measured the RMS noise of each image in an annulus with inner and outer radii of 15 and 30 pixels, respectively, around the phase centre, thereby avoiding target emission. We regard any flux density at least three times the surrounding image RMS as a detection. If the target was detected significantly at this threshold, we repeated this procedure on the four sub-bands of each ALMA band. In all cases, a significant source detection in the full band translated to detections in each of the four sub-bands. All ALMA images and the script to repeat these measurements are included in this paper's Data Reproduction Package. The results of this ALMA analysis are listed in Table \ref{tab:mmresults}.

\subsection{Radio observations}

As part of the ALMA observing campaign, we observed the six ALMA targets with the VLA at 6 and 10 GHz (Program SA110657). The six observations were taken between 15 and 19 October 2024, while the array was in A configuration, with a maximum of 17 day separation between a target's VLA epoch and one of its ALMA band observations. Each target was observed for one hour including overheads, resulting in approximately $13$ minutes on target per observing band. Both bands were setup with 3-bit samplers to yield 4 GHz of bandwidth per band. Different primary and secondary flux calibrators were chosen per target, depending on the source position and the LST of the observation. 

We reduced the observations using \textsc{casa} v.6.6.5.31 \citep{casa2022} to perform data inspection and flagging, calibration, and imaging following standard practices. Specifically, the data were imaged using \textsc{tclean} at both observing frequencies separately, varying the robustness parameter to optimize the trade-off between sensitivity and the effect of side-lobes and imaging artifacts. If a radio counterpart was detected at the known source position, we used \textsc{imfit} to measure its flux density by fitting a 2D Gaussian with full-width-half-maxima and orientation fixed to the synthesized beam. We calculated the flux density error as the RMS variability across a close-by region devoid of sources. For undetected targets, we calculate the RMS across the source position and report the $3\sigma$ upper limit on the flux density. Further details of the observations, including times of observation, optimal robustness parameter, calibrator sources, and flux density limits, are listed in Table \ref{tab:appendix_radio}. 

In two cases, observational issues strongly affect the quality of the data to the level where we do not use their results in the further analysis. In the 6-GHz observation of IGR J18410-0535, large side-lobes and artifacts are present in the image, regardless of imaging parameters and whether the flagging and calibration was performed fully manually or started from the VLA pipeline. These artifacts do not resemble those expected from nearby or diffuse sources. In the 10-GHz observation of 4U 1700-37, good phase calibration solutions using the secondary calibrator were challenging to obtain. Improving the calibration by attempting self-calibration on the target, expected at several hundreds of $\mu$Jy, did not fully resolve these issues. The resulting imaging does reveal a potential source at the target position, but not at $3\sigma$ significance due to the high remaining image noise. The potential counterpart is seen at $207 \pm 87$ $\mu$Jy, which we list in Table \ref{tab:appendix_radio} and include in the SEDs, but do not use in further SED fitting or interpreting the broad-band spectrum.

We finally note that we cross-referenced the position of each target in radio survey data from the Very Large Array Sky Survey \citep{gordon2021}, the MeerKAT Galactic Plane Survey \citep{goedhart2024}, the Rapid ASKAP Continuum Survey \citep{mcconnell2020}, and the ASKAP Evolutionary Map of the Universe (EMU) survey \citep{norris2011}. None of these radio surveys returned a counterpart to the considered system, with unconstraining upper limits in comparison to our VLA and ALMA/NOEMA observations.

\section{Results}
\label{sec:results}

\subsection{Millimeter and radio imaging}

\begin{figure*}
\includegraphics[width=\textwidth]{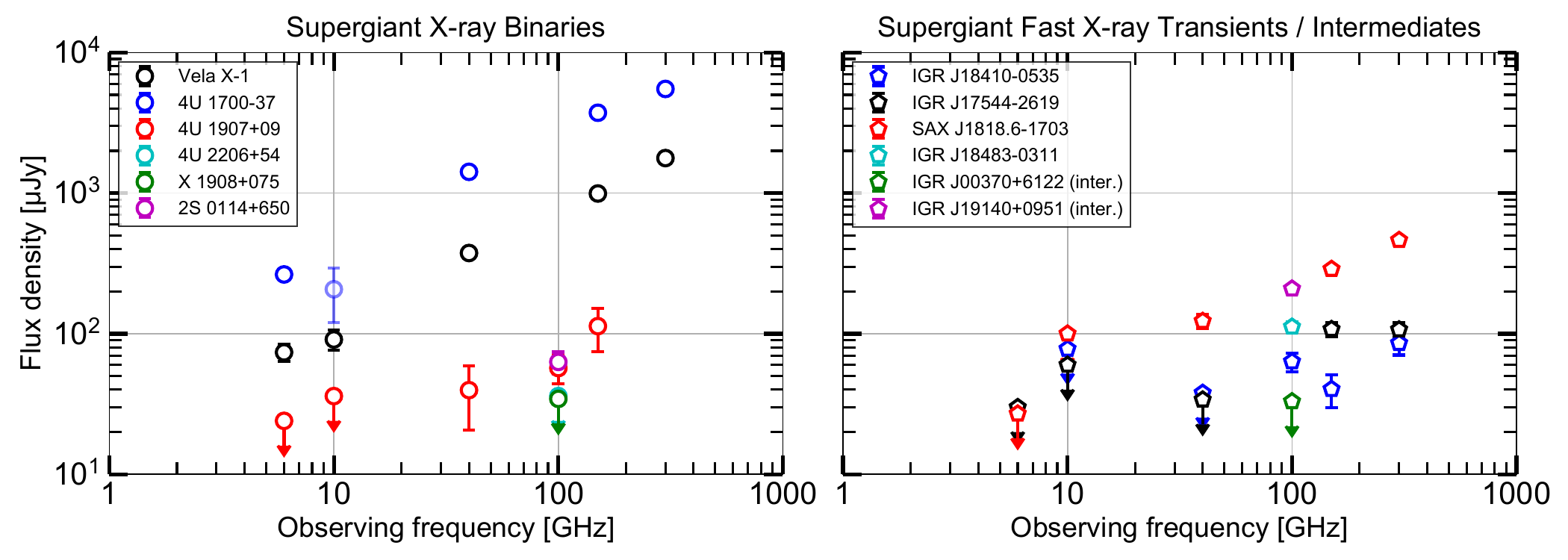}
\caption{The low-frequency spectral energy distributions of all twelve targets, separated into SgXBs (left) and SFXTs / intermediates (right). For each source, we plot all available ALMA and NOEMA data, including those from Van den Eijnden et al. (2023). We also show the coordinated VLA flux density (limits) for the six ALMA targets. At several frequencies, upper limits overlap; in all those cases (particularly the 6/10 GHz bands in the right-hand panel) the overlapping points all represent upper limits. In the left-hand panel at 100 GHz, above the two overlapping limits, two detections are plotted at overlapping flux density.}
\label{fig:SED}
\end{figure*}

\begin{figure*}
\includegraphics[width=\textwidth]{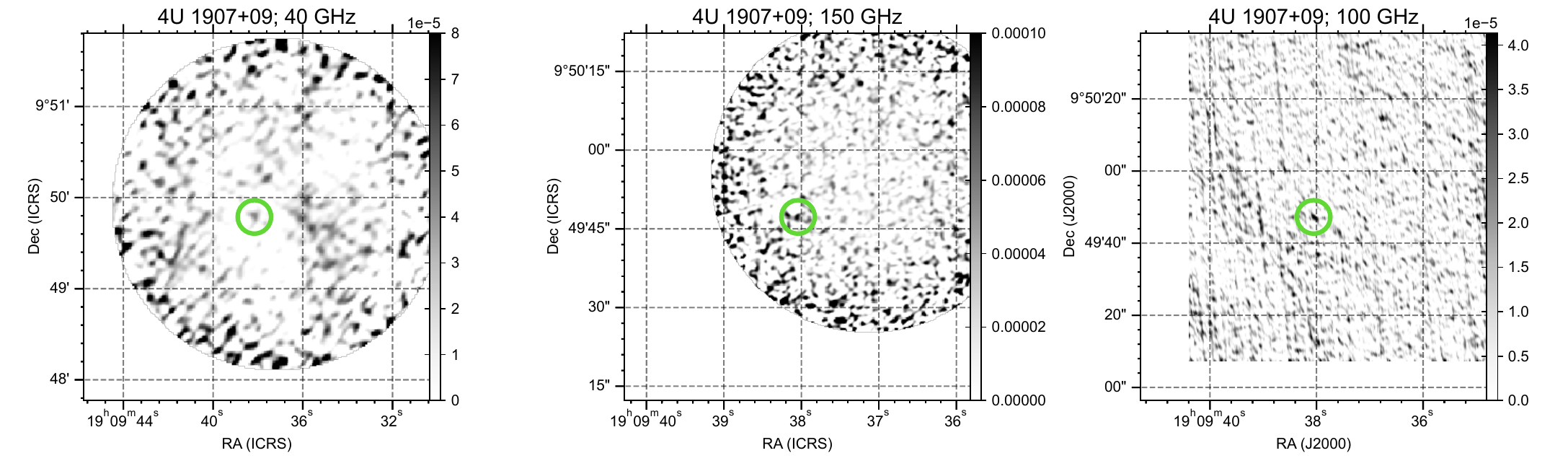}
\caption{The two ALMA images at 40 GHz (left) and 150 GHz (middle), and the NOEMA image at 100 GHz (right) of the field of 4U 1907+09. The green circle in each panel is centred on the position of the SgXB, which is used as the plotting center of each panel. With increasing frequency, the field of view shrinks, causing the pointing centre to shift to the upper right in the ALMA images. While formally not $>3\sigma$ significant in both ALMA images, we measure its flux and report it as a detection due to its consistent presence in all images covering the position; in particular, the NOEMA 100-GHz image shows a millimeter counterpart at $\gtrsim 4\sigma$. The ALMA 300-GHz image, where the source position is not covered, is shown in Figure \ref{fig:all2}.}
\label{fig:4u1907}
\end{figure*}

\begin{figure*}
\includegraphics[width=\textwidth]{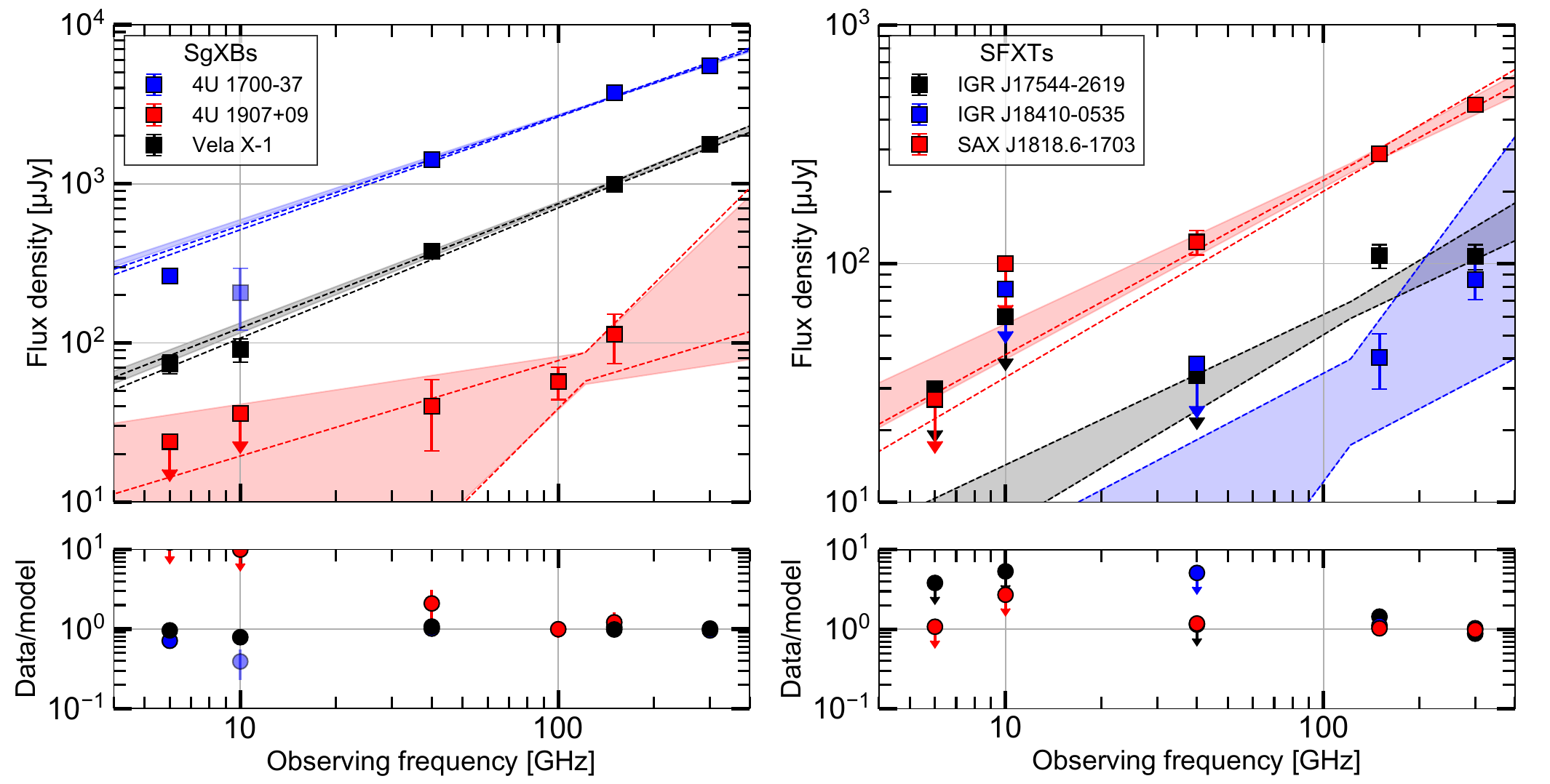}
\caption{Fits of a power-law radio / millimeter spectrum $S_\nu = \xi S_0 (\nu / \nu_0)^{\alpha}$ to the six ALMA targets. The left and right panels show the three SgXBs and SFXTs, respectively, where we stress the difference in vertical extent. The shaded regions and dashed lines indicate the $1\sigma$ confidence regions of the MCMC fits, including only the ALMA data and both the ALMA and VLA data, respectively. Below each spectral fit, the residuals are shown as data divided by model, for the fit to the ALMA and VLA data. Note that we plot the 10-GHz point for 4U 1700-37 but do not use it in the MCMC fit.}
\label{fig:SEDfits}
\end{figure*}

Using the methods described in the previous Section, we assessed the images and measured the flux densities of all ALMA and NOEMA observations. The measured flux densities or $3\sigma$ upper limits are reported in Table \ref{tab:mmresults} and plotted as a function of frequency in Figure~\ref{fig:SED}. Out of the total sample of twelve sources, nine sources are detected in the millimeter band. The first millimeter-detected SFXT, IGR J18410-0535, was re-detected with ALMA, confirming the earlier NOEMA result. In total, five out of the six ALMA targets are significantly detected in two or three bands at the phase centre (see below). Three out of six NOEMA targets were detected at the phase centre. In Table \ref{tab:mmresults}, the results per band are listed; flux densities per sub-band can be found in the Appendix. The Appendix also includes images for all bands and all targets.

It is worth explicitly discussing 4U 1907+09. Due to a slight offset in target coordinates, 4U 1907+09 was not located at the pointing / phase center in both the ALMA and NOEMA observations. This has two important consequences: firstly, it falls just outside the field for the ALMA observation at 300 GHz, where the primary beam is smallest. Secondly, especially for the ALMA observations, the offset from the phase centre increases the local RMS. The two ALMA images covering the position of 4U 1907+09 field (e.g., 40 and 150 GHz) are shown in Figure \ref{fig:4u1907}, as is the NOEMA 100-GHz image. For the two ALMA images, we manually measure the flux density at the offset source position, yielding the flux density levels reported in Table \ref{tab:mmresults}. The local, enhanced RMS yields formally insignificant detections ($<3\sigma$) for the $40$ and $150$ GHz ALMA observations. Yet, we take the presence of a faint point source at the correct position in both ALMA images, as well as the $\gtrsim 4\sigma$ NOEMA detection, as sufficient evidence to claim a source detection. We therefore treat 4U 1907+09 as detected from here on, bringing the total number of detected ALMA targets to six out of six, and detected NOEMA targets to four out of six (considering only newly observed targets from program w23bo). Due to overlap between ALMA and NOEMA targets and including earlier campaigns, nine of twelve total targets are detected in the millimeter band.  

For all targets, we use the observed flux density (limit) and distance to calculate the specific luminosity, defined as $L_\nu = 4\pi F_\nu D^2$ and listed in Table \ref{tab:mmresults}. Ignoring the exact spectral shape by calculating the specific luminosity allows us to compare all targets among each other and with other low-frequency stellar sources. We calculate its error by propagating both flux density and distance errors, explaining the significant errors for some targets through the squared dependence on sometimes poorly constrained distances\footnote{Note that for IGR J18410-0535, we adopt the $6.9\pm1.7$ kpc out of the potential range.}. 

In the ALMA observations, we explored whether any evidence for short-time-scale flaring or narrow emission line features could be seen. These searches were motivated by, respectively, the variability seen in X-rays for both SgXBs and (in particular) SFXTs, and the rest frequency of the H54 $\alpha$ radio recombination line (RRL) at 40.630 GHz. We searched the ALMA observations in each band where the target was detected at $>3\sigma$ significance (e.g., excluding 4U 1907+09) using the \textsc{uvfit} tool in \textsc{casa} and the calibrated measurement set. We measured the target light curves at both a 10 second and 1 minute time resolution, but did not find evidence for flaring or slower flux variations throughout the observations. The poorer uv-coverage and lower sensitivity of NOEMA prevented us from repeating this approach on the NOEMA data. Comparing, however, the NOEMA flux densities of detected targets between individual observing days revealed potential, low-level variations. We will discuss such longer time-scale variability in Section \ref{sec:var}. Finally, we found no evidence for an RRL around 40.630 GHz in the ALMA Band 1 pipeline spectrum of any target.

The results of the VLA radio observations of the six ALMA targets are summarized in Table \ref{tab:appendix_radio}. Significant radio emission was detected from 4U 1700-37 and Vela X-1, the two SgXBs already detected with ATCA by \citet{vandeneijnden2021}. Both sources are detected at lower radio flux densities in our VLA observations, despite the similar observing frequencies to the previous ATCA campaign. The most striking difference is seen for 4U 1700-37, which was detected at approximately twice the flux density by \citet{vandeneijnden2021}. The third SgXB, 4U 1907+09, is not detected at radio frequencies, nor are any of the three SFXTs targeted with the VLA. All detected flux densities or $3\sigma$ upper limits are plotted in the SEDs of Figure \ref{fig:SED}. The radio detections and non-detections appear generally unsurprising given the flux densities and spectral shapes of the ALMA targets, coupled with the typical VLA sensitivities: only the two millimeter-brightest systems are detected in radio. While our VLA observations are, to our knowledge, the deepest constraints on SFXT radio emission to date, this source class remains completely radio undetected. 

\subsection{Low-frequency SED fitting}

We attempt to fit the low-frequency millimeter or radio + millimeter SEDs of our targets with a power law model $S_\nu = \xi S_0 (\nu / \nu_0)^{\alpha}$. For all sources with multiple millimeter band detections (i.e., all ALMA targets), we set $S_0 = 1$ mJy and $\nu_0 = 121.6$ GHz, equal to the mean of the three ALMA centroid frequencies in log space. The choice of model shape reflects the expected power law shape for both a thermal wind (in the regime where the size of the millimeter and radio photosphere exceed the size of the star) and non-thermal jet spectrum. We skip sources with a single-band non-detection (X1908+075, IGR J00370+6122, 4U 2206+54). We note that, for sources with only a single NOEMA band detection, we can use the two NOEMA sub-bands to calculate the spectral index (as the number of bands equals the number of parameters). Here, we find values of $\alpha = -1.2\pm2.1$, $\alpha = -1.0\pm0.7$, and $\alpha = -0.6\pm0.9$ for 2S 0114+650, IGR J19140+0951, and IGR J18483-0311, respectively. Due to the large uncertainties on these values compared to the fits discussed below and, in particular, the small fractional frequency difference ($\sim 0.15$) between the two sub-bands, we do not use these measurements further in our interpretation of the SED shapes. 

For the remaining six (ALMA / VLA) targets, we fit the power law model using an MCMC fitting routine. We take non-detections into account by assigning a zero probability (or, in the fitting routine, a minus infinity log probability) to any model that predicts a flux above a $3\sigma$ upper limit. We assume flat priors on the spectral index $\alpha$ and on the logarithm of the normalization parameter, $\log \xi$. We perform three versions of the SED fit: firstly, we fit the ALMA SEDs using the three full ALMA bands, followed by a fit to the ALMA SEDs using the three times four sub-bands as a consistency check. Subsequently, we also fit the ALMA + VLA SEDs, using the full bands (6, 10, 40, 150, and 300 GHz) for each. Finally, for 4U 1907+09, we only perform a single fit to the ALMA 40 and 150 GHz plus the NOEMA 100 GHz data, given the offset pointing and therefore low signal-to-noise. For each fit, we report the resulting parameter estimates and their $16/84^{\rm th}$ percentile confidence intervals in Table \ref{tab:SEDfits}. We also plot the $1\sigma$ confidence regions of the fitted power law models, for the fits listed in Table \ref{tab:SEDfits}, alongside the data in Figure \ref{fig:SEDfits}. Corner plots from all fits can be found in the Data Reproduction Package of this work. 

\renewcommand{\arraystretch}{1.3}
\setlength{\tabcolsep}{2pt} % Default value: 6pt
\begin{table}
\caption{Results of the power law model MCMC fits, $S_\nu = \xi S_0 (\nu / \nu_0)^{\alpha}$, to the six targets with ALMA + VLA data assuming flat priors on spectral index $\alpha$ and on the logarithmic normalization $\log \xi$. The two different fits differ in whether the VLA radio data was included in the fit. Note that for IGR J17544-2619 and IGR J18410-0535, the fit remains the same, as the VLA data are unconstraining compared to the millimeter band. *For 4U 1907+09, the NOEMA 100-GHz band was included instead of the ALMA 300-GHz band.}
\label{tab:SEDfits}
\begin{tabular}{lcccc}
\hline
Target & \multicolumn{2}{c}{ALMA full bands} & \multicolumn{2}{c}{ALMA full bands + VLA} \\
& $\log \xi$ & $\alpha$ & $\log \xi$ & $\alpha$ \\
\hline
4U 1700-37 & $0.489\pm0.002$ & $0.67\pm0.01$ & $0.484 \pm 0.002$ & $0.70\pm0.01$ \\
Vela X-1 & $-0.06 \pm 0.01$ & $0.78 \pm 0.02$ & $-0.07 \pm 0.01$ & $0.80 \pm 0.02$ \\
4U 1907+09* & $-1.15^{+0.09}_{-0.11}$ & $1.0^{+0.9}_{-0.7}$ & $-1.14^{+0.08}_{-0.10}$ & $1.2^{+0.8}_{-0.6}$ \\
IGR J17544-2619 & $-1.19^{+0.03}_{-0.04}$ & $0.70^{+0.10}_{-0.07}$ & $-1.19^{+0.03}_{-0.04}$ & $0.70^{+0.10}_{-0.07}$ \\
IGR J18410-0535 & $-1.55^{+0.15}_{-0.21}$ & $1.2^{+0.6}_{-0.5}$ & $-1.55^{+0.15}_{-0.21}$ & $1.2^{+0.6}_{-0.5}$ \\
SAX J1818.6-1703 & $-0.60\pm0.02$ & $0.67\pm0.05$ & $-0.62\pm0.01$ & $0.75^{+0.03}_{-0.02}$ \\ \hline
\end{tabular}\\
\end{table}
\renewcommand{\arraystretch}{1.0}

As expected from the SED shown in Figure \ref{fig:SED}, we find inverted spectral indices for all six targets. Particularly, the lowest spectral index taking into account the $1\sigma$ uncertainties is found for 4U 1907+09: $\alpha = 1.0^{+0.9}_{-0.7}$, from the millimeter data only. Regardless of the inclusion or exclusion of radio data, the spectral indices are typically slightly higher than expected in the theoretical \citet{wright1975} and \citet{panagia1975} prediction of $\alpha = 0.6$: for the most accurately measured spectral indices, we find values in the approximate range $\alpha \approx 0.6 - 0.8$. Such deviations may be expected in a thermal stellar wind by relaxing assumptions around its velocity structure and acceleration profile: a wind that is still accelerating at the typical emission radii probed by the radio and millimeter bands will show a spectral index in excess of $\alpha = 0.6$ \citep[see also][]{erba2022}. 

Comparing our different fits, we find that only for IGR J17544-2619 and IGR J18410-0535, differences in the fit parameters are found when changing the full ALMA band data to the ALMA sub-bands. In those cases, the two parameters remain poorly constrained in comparison to the full ALMA bands. This change is to be expected: for these two weakest millimeter sources, the sub-bands are at such low signal-to-noise that they remain poorly constraining on the spectral index and flux density normalization. This effect is particularly striking for IGR J17544-2619, where the 40-GHz upper limit is strongly constraining on the power law shape when using the full ALMA bands; however, in the sub-bands, with $\sim 2$ times higher RMS uncertainties, these non-detections become consistent with a much broader range of spectral indices. Discussing IGR J17544-2619, we should also note that its full-ALMA-band SED does not appear to follow a single power law well; in fact, we may be observing the transition frequency, where the photosphere becomes smaller than the stellar radius, and the resulting flattening of the spectrum between the 40 and 100 GHz bands.

Comparing the fits with and without VLA data, only minor differences are found: slight changes, just beyond the $1\sigma$ confidence levels, are seen for the three brightest sources (4U 1700-37, Vela X-1, SAX J1818.6-1703). These result, as is visible in Figure \ref{fig:SEDfits}, in slightly shifted but similar confidence regions for the model. In all three cases, these changes are driven by the lower radio flux densities or limits observed than extrapolated from the millimeter-only fit. The time difference between the observations may be responsible for these discrepancies, in particular for 4U 1700-37, where the deviation at 6 GHz is most significant -- we will further discuss this scenario in Section \ref{sec:var}.

\section{Discussion}
\label{sec:discussion}

In this work, we present a systematic study of the millimeter and centimeter radio emission of twelve neutron star HMXBs. Combining existing and new ALMA and NOEMA observations, nine out of twelve targets are detected at one or more millimeter frequencies. These observations may shed light on the properties of the donor star's wind in neutron star HMXBs, but may also be affected by non-thermal processes contributing at low frequencies. In this Discussion, we will investigate the origin of the detected low-frequency emission, before turning to  inferences on the stellar wind properties. We will subsequently discuss additional constraints on the winds for systems that power a bow shock in the ISM, and will briefly discuss evidence for low-frequency variability in our sample.

\subsection{The origin of radio -- millimeter emission in SgXBs and SFXTs: stellar winds and/or jets?}

To understand the origin of low-frequency emission in neutron star HMXBs, the evident comparison sources are those that show some similarity in their constituents: isolated OB supergiants, for instance, or neutron star X-ray binaries with a different type of donor star (Be/X-ray binaries, low-mass X-ray binaries). In these comparison source classes, respectively, the radio emission originates as thermal emission from the stellar wind, or as non-thermal emission from relativistic jets. We will therefore consider these two options in this Section. We note that this approach treats these thermal and non-thermal processes as independent: it thereby ignores potential interactions between jets and stellar winds \citep[e.g.,][]{zdziarski2012,lopez2022} or non-thermal emission from winds \citep[(such as shock in colliding wind binaries;][]{dougherty2003,bloot2022}. In our thermal wind discussion, we do not include Wolf-Rayet stars, as none of the donor stars in our systems are of this stellar type.

For the ionized, outer winds of massive stars, analytical models predict inverted radio to millimeter spectra ($\alpha > 0$), with a canonical spectral index prediction of $\alpha = 0.6$ \citep{wright1975,panagia1975}. Observations of OB supergiants, as well as these analytical models, suggest specific radio luminosities in the range of $10^{17}-10^{19}$ erg/s/Hz, with the upper end similar to that in extreme cases such as Wolf-Rayet stars \citep{gudel2002}. In the radio band, OB supergiant winds are therefore expected at $\sim 10^{26}-10^{28}$ erg/s. 

For relativistic jets, most spectral constraints exist for black hole X-ray binaries and neutron stars with low-mass donors. Steady jets in both types of systems are typically seen with flat or slightly inverted spectra \citep[$\alpha \gtrsim 0$;][]{fender2001}, but most typically not as inverted as the stellar wind regime. Such spectra are also seen, when multi-band observations are available, for the jets launched in Be/X-ray binaries \citep[with the exception of an optically thin radio spectrum in the super-Eddington state of Swift J0243.6+6124;][]{vandeneijnden2018}. The jet's luminosity, for strongly-magnetized neutron stars \citep[like those in HMXBs; $B\gtrsim10^{12}$ G;][]{staubert2019}, is best measured in Be/X-ray binaries as well, given the weakness of the spherical stellar wind that complicates our interpretation here for SgXBs and SFXTs: interpolating the Be/X-ray binary jets to lower accretion rates, we expect a radio luminosity of the order $\sim 10^{26}$ erg/s at an accretion rate of $\sim 1$\% Eddington, typical for a SgXB \citep{vandeneijnden2024_lsv4417}. 

Several complications in these jet expectations remain, however: firstly, the interaction with the wind may brighten the jet, as discussed in \citet{vandeneijnden2024_lsv4417} for a so-called persistent Be/X-ray binary. Secondly, the Be/X-ray binary Swift J0243.6+6124 deviated from the extrapolation used above for Be/X-ray binary jet luminosities during brief X-ray re-brightenings. Therefore, the prediction above may be underestimated by two orders of magnitude, although the origin of this deviation is not well understood. On the other hand, the jets in Be/X-ray binaries are launched from an accretion disk, which is not likely to be present in most SgXBs and SFXTs \citep[although see][]{elmellah2018}; depending on the jet launch model, that may complicate the formation of the jet in the first place \citep{massi2008,parfrey2016,das2022,das2024}.

Given these two possible contributing processes (winds and jets) and the complications in predicting the jet scenario exactly, previous radio studies remained ambiguous on the emission's origin. \citet{pestalozzi2009}, for instance, report radio monitoring of the neutron star + hypergiant system GX 301-2 / BP Cru, reporting variability in radio brightness and spectrum. Whether this is driven only by the stellar wind or includes an additional non-thermal component at certain orbital phases remains unconfirmed in their data, also due to the lower sensitivity in the pre-CABB ATCA data \citep{CABB}. In \citet{vandeneijnden2021}, the radio counterpart of five SgXBs is reported. However, the uncertainties on the spectral index, measured over the small range from $5.5$ to $9$ GHz, are too large to unambiguously associate this emission with either a wind, a jet, or their interaction.  

The millimeter observations presented in this work help answer this question in more detail, particularly using the combined VLA and ALMA constraints for six targets. There, the frequency range from 6 to 300 GHz reveals strongly inverted low-frequency spectra, with the spectral index typically observed to lie in the range $\alpha \approx 0.6-0.8$\footnote{Interestingly, a similar spectral slope with $\alpha = 0.71\pm0.02$ was found in recent \textit{JWST} MIRI observations of the black hole X-ray binary A0620-00 in its quiescent state, interpreted as thermal free-free emission from a strong and warm ($T_e \approx 10^4$ K) wind outflow from the accretion disc \citep{2025arXiv250523918Z}.}. The spectral indices measured from NOEMA sub-bands do not provide similar constraints, given their significant uncertainties and small fractional frequency differences. The specific luminosities measured in the millimeter bands range in the order a few times $10^{17}$ to a few times $10^{18}$ erg/s/Hz at the lowest millimeter frequencies (40 or 100 GHz), which is comparable to the typical levels of early-type single and binary stars. The 300 GHz millimeter luminosity in 4U 1700-37, for instance, is comparable to the brightest 300 GHz luminosity seen from the jet in the neutron star low-mass X-ray binary Aql X-1 \citep{diaz2018}; a jet interpretation in this SgXB would be at odds with the much fainter radio jets seen in Be/X-ray binaries than in neutron star low-mass X-ray binaries. 

Given the new evidence from our radio + millimeter spectra, we conclude that these low-frequency SEDs are dominated by the thermal free-free emission from the stellar wind. For six sources, we do not have a spectrum but merely a single band NOEMA measurement (as the sub-bands are separated by a small fractional difference, we opt not to put much weight on the measured in-band spectral index). As the typical specific luminosities of those targets, when detected, are similar to the VLA+ALMA targets, we deem it most likely that their emission is similarly dominated by the stellar wind; no systematic difference in target selection was applied between the two sub-samples, beyond declination constraints per telescope. 

In this stellar-wind free-free scenario, the observed spectral indices deviate from the canonically expected $\alpha=0.6$. This deviation may indicate that the wind has not reached its terminal velocity yet, in the regions probed by the centimeter/millimeter bands. Following \citet{panagia1975}, the spectral index $\alpha$ depends on the wind's density profile $n_e(r) \propto r^{-p}$ as $\alpha = 2 - 4.2/(2p-1)$. The canonical spectral index corresponds to $p=2$, while the observed range of $\alpha = 0.6-0.8$ implies $p=2-2.25$. Under mass conservation, the density profile scales as $n_e(r) \propto r^{-2}v(r)^{-1}$. Combined, therefore, $p>2$ implies that $v(r) \neq v_\infty$ at the considered radius, but instead still increases with radius -- inn other words, still accelerates. We note that wind velocity profiles are usually modeled as $\beta$-profiles \citep{puls2008}, which result in more complex radial density profiles than assumed by \citet{panagia1975}; however, qualitatively, the same argument holds in that case.

In none of the six VLA + ALMA target can we see evidence for a radio jet dominating at lower frequencies, only to be taken over by a more inverted stellar wind spectrum in the millimeter band. To see to what degree that may be expected in future observations, we can perform an order of magnitude comparison of the expected luminosity of the stellar wind and jet in a neutron star HMXB. For this, we assume a stellar wind mass loss rate and velocity, from which we calculate the analytical thermal wind spectrum. Further assuming a $1.4$ $M_\odot$ neutron star, a donor star mass, and an orbital period, we can calculate the orbital velocity $v_{\rm orb}$ assuming a circular orbit and the ratio $q$ of neutron star to total mass. Following \citet{tejeda2025}, we can then estimate the wind capture fraction as $\eta_{\rm cap} = (q / (1 + (v_\infty / v_{\rm orb}))^2$, which gives an X-ray luminosity of $L_X = 0.1 \eta_{\rm cap} \dot{M}_{\rm w} c^2$. Finally, we estimate the expected jet luminosity from the best-fit $L_X-L_R$ relation for Be/X-ray binaries reported in \citet{vandeneijnden2022}. We emphasize that each of these steps in the calculation of the jet luminosity comes with statistical and systematic uncertainties, and therefore this comparison should be interpreted only at the order of magnitude level. 

\begin{figure}
\includegraphics[width=\columnwidth]{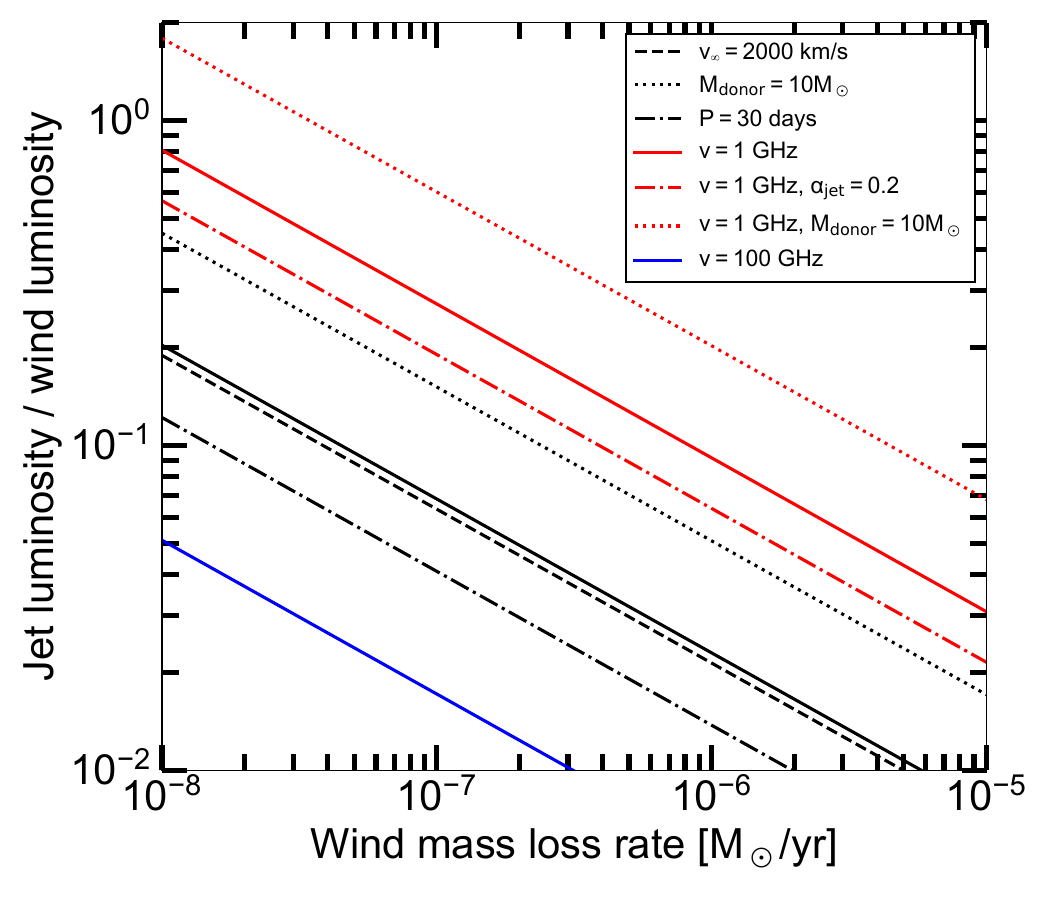}
 \caption{An order of magnitude comparison of the jet and wind luminosity assuming wind capture in a circular orbit and the measured Be/X-ray binary coupling between accretion rate and jet power. The legend indicates deviations from the standard input, shown as the black line ($v_\infty = 1000$ km/s, $\nu = 10$ GHz, $M_{\rm donor} = 20$ $M_\odot$, $P_{\rm orb} = 10$ days, $\alpha_{\rm jet} = 0$). Despite the assumptions underlying this calculation, the jet is unlikely to be visible above $1$ GHz unless interactions with the wind brighten it significantly.}
\label{fig:jet_vs_wind}
\end{figure}

In Figure \ref{fig:jet_vs_wind}, we show the ratio of jet and wind luminosity as a function of wind mass loss rate, varying a number of the input parameters: the observing frequency, wind terminal velocity, jet spectral index, orbital period, and donor star mass. The black line shows the baseline case ($v_\infty = 1000$ km/s, $\nu = 10$ GHz, $M_{\rm donor} = 20$ $M_\odot$, $P_{\rm orb} = 10$ days, $\alpha_{\rm jet} = 0$), while the legend shows the varied parameters for each of the other curves. From this order-of-magnitude comparison, we find that beyond $1$ GHz frequencies, the jet is unlikely to dominate in the spectrum of SgXBs; even at low frequencies, that only occurs in specific circumstances. While these exact estimates will be affected by a range of complications -- e.g., the wind acceleration profile, orbital eccentricity, uncertainties in the accretion rate -- jet power relation -- two conclusions can be drawn: if a combined spectrum, dominated by a jet at low frequencies, is observed in a SgXB in future campaigns, the jet's luminosity is likely enhanced from interactions with the wind material. Secondly, the Be/X-ray binary jets dominate due to a combination of weak stellar wind and disk-accretion instead of wind-accretion, de-coupling the mass accretion rate from the stellar wind parameters. 

\subsection{Stellar wind properties in SgXBs and SFXTs}

\subsubsection{How do the millimeter properties of stellar winds compare between SFXTs and SgXBs?}

The difference in X-ray behavior between SgXBs and SFXTs may be related to the properties of the donor star and its wind, the orbital parameters, the neutron star, and combinations thereof. As discussed in the reviews by \citet{Sidoli2017} and \citet{Kretschmar2019}, none of these three options individually offer an obvious explanation: the sample of known SFXTs and SgXBs shows significant overlap in their observables, such as donor star types, orbital periods, and neutron star spin periods. Differences in clumping structure in the winds are unlikely to account for the flaring behaviour in SFXTs, given the unrealistically large clump masses needed to power a typical X-ray flare. Through a detailed and consistent comparison between two systems, the SFXT IGR J17544-2619 and SgXB Vela X-1, \citet{gimenez2016} suggested that the stellar wind in the SFXT may be significantly faster -- leading to a lower capture rate and average X-ray luminosity. Similar arguments for the overall sample were made by, e.g., \citet{negueruela2019}, arguing that the stellar types in SFXTs may be earlier and that therefore their winds may be faster. Alternatively, \citet{hubrig2019} discussed the role of the stellar wind's magnetic field in regulating the wind accretion process.

\begin{figure*}
\includegraphics[width=\textwidth]{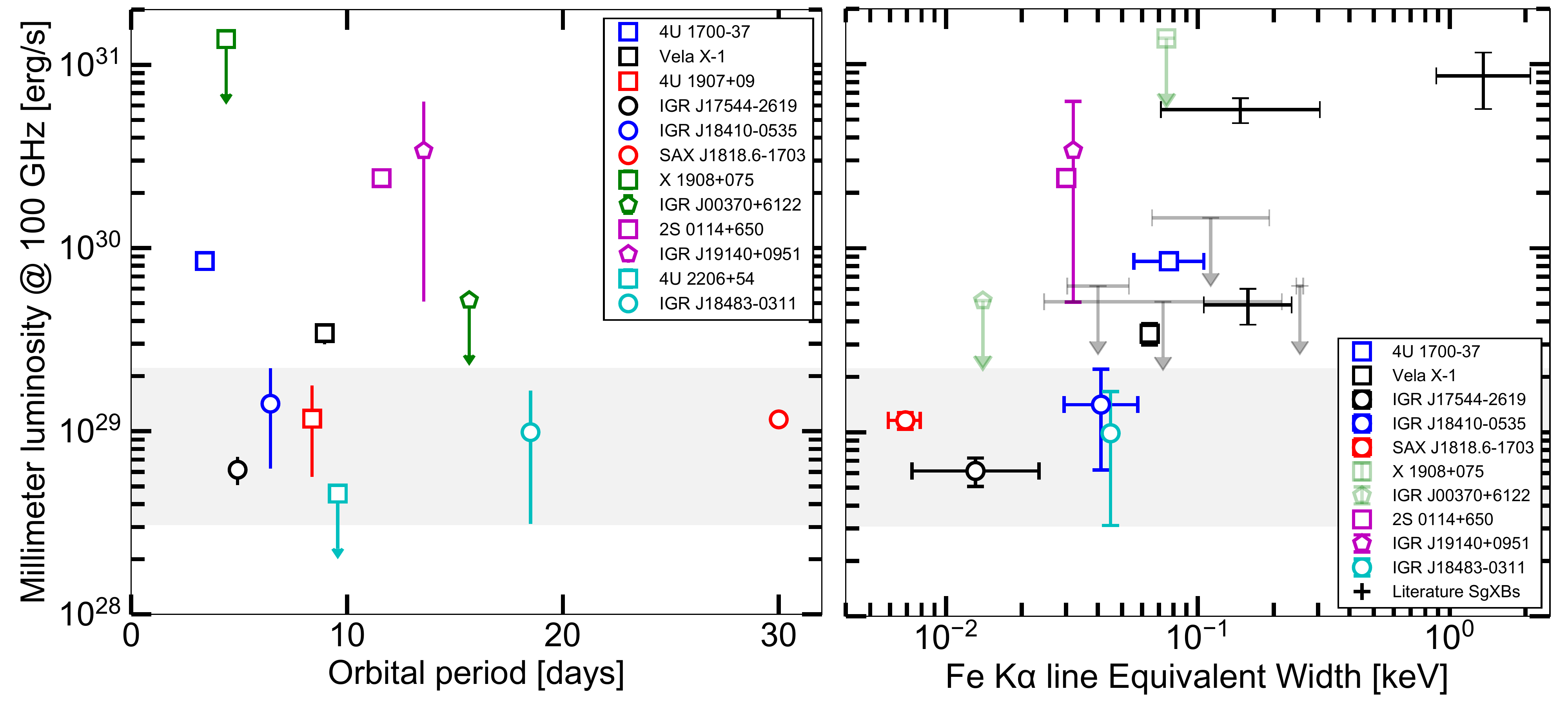}
 \caption{\textit{Left:} The 100-GHz millimeter luminosities versus orbital period. SgXBs are shown as squares; SFXTs as circles; intermediate sources as pentagons. The uncertainties on the millimeter luminosity include both flux density and distance uncertaintes. \textit{Right:} the millimeter luminosity plotted versus the Equivalent Width of the Fe K$\alpha$ line at $6.4$ keV. Archival points are shown based on \citet{vandeneijnden2021}, under the assumption that these extrapolated radio luminosities are dominated by stellar wind emission. In both panels, the grey band indicated the range covered by SFXTs.}
\label{fig:Lmm_vs_P}
\end{figure*}

Whether any such systematic wind differences indeed exist and drive the behavioral dichotomy between SFXTs and SgXBs, remains, at the moment, unconfirmed. As we conclude that the millimeter emission of both types of systems is dominated by thermal wind emission, our data particularly allows us to investigate the case of different donor star properties. With this in mind, we show the re-scaled 100-GHz luminosity versus the orbital period of our targets in the left panel of Figure \ref{fig:Lmm_vs_P} (using the observed spectral index, or otherwise assuming $\alpha = 0.6$). Comparing the millimeter luminosity allows us to understand the role of source distance and removes the differences due to the different NOEMA and ALMA observing frequencies. The four SFXTs in our sample (the circles) show a strikingly small range ($\sim 3.3$) in millimeter luminosity, between $6\times10^{28}$ erg/s and $2\times10^{29}$ erg/s. The set of all six persistently accreting systems, on the other hand, trace a wider luminosity range (a factor $\gtrsim 50$), between $<4.6\times10^{28}$ erg/s and $2.4\times10^{30}$ erg/s. The intermediate system IGR J19140+0951 sits among the brightest millimeter sources.

A more detailed look at these millimeter luminosities -- taking into account the properties of the individual SgXBs -- reveals a potential difference between the SFXTs and the \textit{classical} SgXBs. Firstly, 4U 2206+54 and 4U 1907+09 overlap with the millimeter luminosity range of the SFXTs. As noted earlier, while 4U 2206+54 is grouped with the SgXBs given its persistently accreting nature, its donor is not a supergiant but a O9.5V star \citep{blay2006}. 4U 1907+09 does host a supergiant donor, but displays dipping behaviour in X-rays that has been suggested to be similar to the X-ray flaring in SFXTs, which would make it a link between the two sub-classes \citep{doroshenko2012}. On the other hand, the three millimeter-detected SgXBs with higher 100-GHz luminosities are bona fide SgXBs, in particular the prototypical SgXBs Vela X-1 and 4U 1700-37. We may therefore interpret our results as showing a systematically higher millimeter luminosity in prototypical SgXBs than in SFXTs. 

To further investigate and substantiate this claim, we also perform a comparison between millimeter luminosities and the equivalent width (EW) of the 6.4 keV Fe K$\alpha$ line. This emission line, in HMXBs, originates from the fluorescence in the stellar wind material of X-rays emitted by the accretion flow close to the neutron star. A higher density of wind material therefore leads to a larger EW. The EW is known to correlate with the absorption column density in X-rays \citep{pradhan2018}, which is expected for such higher densities of fluorescing material \citep[see also][]{torrejon2010b,gimenezgarcia2015}. Strikingly, \citet{pradhan2018} report that SgXBs trace a significantly larger range in EW and absorption column than SFXTs, where the SFXTs only overlap with the low end of the EW and absorption column measurements in SgXBs -- behaviour somewhat reminiscent of the millimeter luminosities discussed above. 

For each target in our sample, we therefore take all measurements of the Fe K$\alpha$ EW from \citet{pradhan2018}, regardless of X-ray properties during the observation; instead, we calculate a weighted average of the EW for our comparison. We obtain EW measurements for all targets except 4U 2206+54 and 4U 1907+09 -- precisely the two sources overlapping in millimeter luminosity with the SFXTs and with complication in their SgXB classification. In addition, we take the SgXBs included in the radio compilation of \citet{vandeneijnden2021} and perform a similar cross-match. This literature addition adds the SgXBs OAO 1657-415, 4U 1538-522, EXO 1722-363, IGR J16207-5129, IGR J16318-4848, and IGR J16320-4751 to the sample. Finally, we add GX 301-2 (BP Cru), based on \citet{pestalozzi2009}. These literature-based SgXBs were all observed with ATCA at 5.5 and 9 GHz. The first four systems in the list are not detected in either band; we therefore take their 9-GHz $3\sigma$ upper limit and extrapolate to 100-GHz assuming a standard spectral index $\alpha = 0.6$. For IGR J16320-4751 and GX 301-2, we assume the same spectral shape to extrapolate their 5.5 GHz and highest-significance 9-GHz detections, respectively. Finally, for IGR J16318-4848, we extrapolate its combined 5.5 and 9 GHz detections to higher frequency through a Monte-Carlo simulation. We stress that these extrapolations, and comparison with our millimeter-observed sample, implicitly assume their low-frequency SED is stellar-wind dominated. 

The resulting comparison of millimeter luminosity and EW is shown in the right panel of Figure \ref{fig:Lmm_vs_P}, where millimeter limits are shown semi-transparently and literature-based extrapolations only with errorbars. We observe a confirmation that the SFXTs are millimeter-underluminous compared to the SgXBs with EW measurements. As the SgXBs included in this comparison generally show higher EWs than the SFXTs, we find that the SgXBs typically inhabit a different region of the EW-$L_{\rm mm}$ parameter space than SFXTs. We do note that the (extrapolated) millimeter upper limits for non-detected SgXBs are not constraining, currently, compared to the SFXT luminosity range; deeper observations can therefore further confirm this suggested dichotomy. 

How can we physically interpret this apparent difference between the compared SgXB and SFXTs? Discussing the EW (and its correlation with absorption column), \citet{pradhan2018} discuss two interplaying scenarios: firstly, an overall faster wind in SFXTs would lead to a lower density at the neutron star's position and lower rates of gas capture. Secondly, bright accretion-driven X-ray emission slows down the wind in the vicinity of the neutron star by ionizing the wind material and inhibiting the wind acceleration mechanism, regardless of the overall wind velocity. Such ionizing emission is however, typically, significantly weaker in SFXTs compared to SgXBs. The lower wind density leads to a lower EW if the Fe K$\alpha$ emission predominantly originates from the regions close to the neutron star. 

The millimeter luminosity of our targets provides a large-scale view of the stellar wind in comparison to the small-scale constraints from the EW. Interestingly, these large-scale and small-scale results present a similar picture: the thermal millimeter emission is a probe of the wind density on scales beyond the binary orbit, which decreases for a higher global wind velocity. Barring systematic differences in mass loss rate between the SFXTs and SgXBs, a systematically lower wind velocity may account for the brighter millimeter emission from large scales in SgXBs \textit{and} a higher Fe K$\alpha$ EW from scales close to the neutron star. 

\subsubsection{How do the millimeter properties of stellar winds compare with literature estimates?}
\label{sec:windprops}

\begin{figure}
\includegraphics[width=\columnwidth]{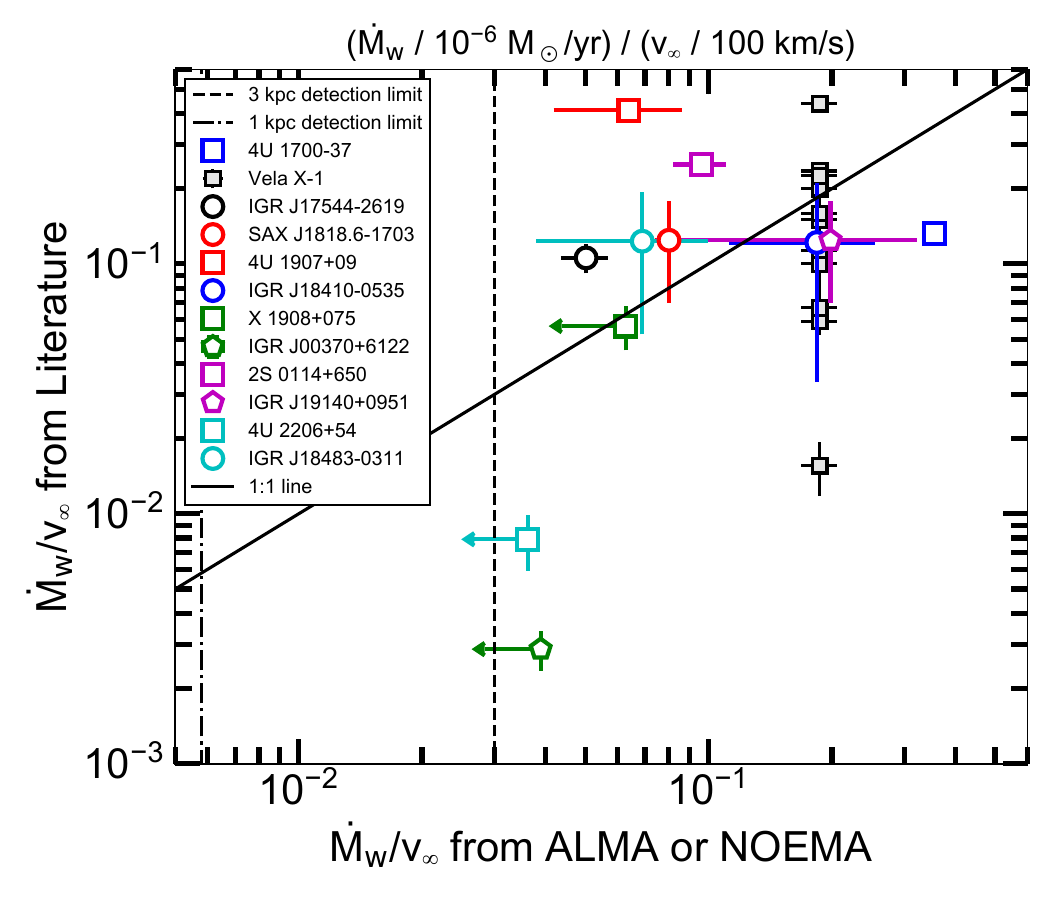}
 \caption{A comparison of the constraints on $\dot{M}_{\rm w} / v_\infty$ from millimeter observations and from literature measurements. Vela X-1 shown as a vertical band reflecting the range of literature measurements; for all other sources, a single literature value is plotted. Squares indicate SgXBs, circles SFXTs and intermediates. The four SFXTs/intermediates sharing the same approximate literature value were plotted using estimates given their stellar type, explaining their similar vertical level.}
\label{fig:literaturecomp}
\end{figure}

Beyond comparing the stellar wind properties as probed in the low-frequency SED between source classes, we can also compare with literature estimates. For that purpose, the observed low-frequency spectrum should be converted to physical properties that can be compared to independent measurements. Specifically, we will use the millimeter luminosity as a proxy for the ratio of mass-loss rate and wind terminal velocity, following the analytical derivations by \citet{wright1975,panagia1975} introduced priorly. Before doing so, we reiterate that this classic, analytical understanding of the low-frequency emission does not fully explain the observed SEDs of our targets: most strikingly, we observe spectral indices that are inverted, but differ from the expected $\alpha = 0.6$, possibly reflecting that the wind may not have reached its terminal velocity in the dominant millimeter emission regions.

The normalization of the thermal wind SED with spectral index $\alpha = 0.6$, in the analytical models, depends on the temperature of the wind, its composition, the distance to the target, and, crucially, the wind density. Under the assumption of a smooth and spherically symmetric wind, mass conservation, and a constant velocity (which will be re-evaluated in a companion work in preparation, further modeling the SEDs presented here) the radial density profile follows from the mass loss rate and velocity. As a result, analytically, the thermal wind flux at radio or millimeter frequency $\nu$ is expected to scale as \citep[see also][for the exact numerical constants]{gudel2002}

\begin{equation}
    S_\nu \propto \left(\frac{\dot{M}_{\rm wind}}{v_\infty}\right)^{4/3} \nu^{6/10} .
    \label{eq:analytic}
\end{equation}

\noindent Therefore, assuming similar compositions and wind temperatures (although the flux density depends only weakly on temperature), we can use our millimeter measurements as a proxy for $\dot{M}_{\rm wind} / v_\infty$. Through alternative measurements, this ratio can similarly be constrained, particularly allowing for a comparison of our millimeter results with probes of the inner stellar wind.

For this purpose, we compiled literature measurements of the stellar wind mass loss rates and velocities for all our targets, obtained from optical / UV spectroscopy and therefore probing the inner wind regions. For eight out of twelve sources, at least one such measurement was available, as summarized in Table \ref{tab:appendix_literaturewinds}. For the other four targets (IGR J18410-0535, SAX J1818.6-1703, IGR J19140+0951, and IGR J18483-0311), we instead match their stellar type to the closest comparison(s) in \citet{crowther2006} and \citet{markova2008}. We then use the reported wind properties of those closest matches to calculate a typically expected $\dot{M}_{\rm wind} / v_\infty$, with an uncertainty representing the spread in values amongst these matched sources. We note that, for all twelve targets, the statistical uncertainties on literature mass loss rates and velocities may be smaller than those from the millimeter constraints. However, systematic uncertainties in the underlying method likely introduce more significant errors $\dot{M}_{\rm wind} / v_\infty$ than we report here. 

In Figure \ref{fig:literaturecomp}, we show the comparison between $\dot{M}_{\rm wind} / v_\infty$ based on the millimeter emission at 100 GHz (extrapolated from 150 GHz for ALMA targets using the measured ALMA-based $\alpha$) and based on literature values using optical or UV spectroscopy -- in other words, a comparison between outer and inner wind diagnostics. All sources are shown with the same markers as in other plots, except Vela X-1: the smaller grey-filled black squares indicates the range of different measurements of $\dot{M}_{\rm wind} / v_\infty$ from the large number of archival studies. Overall, we do not observe that the sample follows the potentially expected one-to-one relation. Instead, we can identify several types of behavior in this diagram: firstly, the four targets matched to earlier studies of other B-type supergiants align in their literature value, and match the one-to-one line within uncertainties. Secondly, four systems lie significantly away from the $1:1$ relation: 4U 1700-37 below the relation, versus 4U 1907+09, 2S 0114+650, and IGR J17544-2619, above. Finally, the three systems that are not detected in millimeter, the non-detection is consistent with their inner wind constraints in all cases. 

The three systems lying above the relation may be accounted for by radially-dependent clumping \citep[][]{sundqvist2018}: clumping can lead to overestimates of the mass-loss rate at a given velocity, increasing $\dot{M}_{\rm wind} / v_\infty$. If the degree of clumping decreases radially towards the outer wind, systems may then be pushed above the 1:1 relation in this figure. Indeed, for two of these three sources, 4U 1907+09 and 2S 0114+650, clumping was not taken into account when estimating the stellar wind parameters \citep{cox2005,reig1996}. In the third, IGR J17544-2619, \citet{gimenez2016} did include clumping in their modeling of the IR, optical, and UV spectra. However, the degree of clumping could not be determined self-consistently and the authors note that more work is needed; therefore, it may be possible that the inferred wind parameters are affected by an underestimate of the degree of clumping. This work analysed Vela X-1 through the same approach, managing to determine the clumping more confidently. This is reflected by the resulting ratio of $0.09^{+0.06}_{-0.04}$ in our units. This measurement for Vela X-1 is close to the millimeter value of $0.19 \pm 0.02$, albeit slightly lower: opposite to sources where clumping affects the measurement. 

Vela X-1 is, generally, a special case in this comparison due to its range of archival estimates. The biggest outliers, both above and below the millimeter constraint, were obtained in early estimates \citep{conti1978,dupree1980,mccray1984,sadakane1985,sako1999} that also typically constrain either mass loss rate or velocity after assuming the other from earlier works. More recent works align more closely with the millimeter measurements, reporting ratios in the range $0.1-0.2$ based on X-ray line measurements \citet{watanabe2006} or wind modeling with the explicit inclusion of X-ray irradiation \citep{krticka2012,sander2018}. While these measurements may be below the millimeter constraint, they do not deviate as significantly as 4U 1700-37. In those cases, variability in the wind may play a role: indeed, as we will discuss in Section \ref{sec:var}, both Vela X-1 and 4U 1700-37 show evidence for slow variability in their low-frequency measurements. That would imply, however, that any effects of variability in other sources are smaller than the effects of clumping and measurement uncertainties. 

From this literature comparison, we find that for sources with a dedicated UV/optical wind characterization \textit{and} a millimeter detection, these complementary wind constraints do not agree perfectly. Wind density structure and variability may underlie these differences, highlighting the case for joint analyses of these multi-wavelength wind constraints and for coordinated multi-wavelength observations.

\setlength{\tabcolsep}{12pt} % Default value: 6pt
\begin{table*}
\caption{The literature constraints on mass-loss rate and terminal velocity, with uncertainties taken from the literature when mentioned in the original reference (final column). For the four targets without literature estimates, we show the mean and standard deviation in measurements of $\dot{M}_{\rm w} / v_\infty$ in stars of similar types (final column) from \citet{crowther2006} and \citet{markova2008}. Here, $\dot{M}_{\rm w} / v_\infty$ is calculated after re-scaling the mass-loss rate to $10^{-6}$ $M_{\odot}$/yr and the terminal velocity of $100$ km/s. *For Vela X-1, we show the range in mass-loss rates and terminal velocities as reported in the literature and compiled by \citet{kretschmar2021}. Original values were reported in \citet{conti1978,dupree1980,mccray1984,sadakane1985,sato1986,sako1999,watanabe2006,krticka2012,falanga2015,manousakis2015,gimenez2016,sander2018}.}
\label{tab:appendix_literaturewinds}
\begin{tabular}{lllll}
\hline
Name & Mass-loss rate [$M_{\odot}$/yr] & Terminal velocity [km/s] & Method & Reference \\ \hline
4U 1700-37 & $\left(2.5^{+2.5}_{-1.3}\right)\times10^{-6}$ & $1900\pm100$ & UV/optical & \citet{hainich2020} \\
Vela X-1 & $(0.265 - 2.4)\times10^{-6}$ & $532-1700$ & UV/optical/X-rays & \citet{kretschmar2021}* \\
4U 1907+09 & $7\times10^{-6}$ & $1700\pm30$ & UV/optical & \citet{cox2005} \\
IGR J17544-2619 & $\left(1.6^{+0.9}_{-0.6}\right)\times10^{-6}$ & $1500\pm200$ & UV/optical/IR & \citet{gimenez2016} \\
X1908+075 & $2.8\times10^{-7}$ & $500\pm100$ & IR & \citet{martinez2015} \\
IGR J00370+6122 & $\left(3.2^{+1.8}_{-1.2}\right)\times10^{-8}$ & $(1100\pm200)$ & UV/optical & \citet{hainich2020} \\
2S 0114+650 & $3\times10^{-6}$ & 1200 & Optical/IR & \citet{reig1996} \\
4U 2206+54 & $\left(3.2^{+3.1}_{-1.7}\right)\times10^{-8}$ & $400\pm100$ & UV/optical & \citet{hainich2020} \\ \hline
& Mean $\dot{M}_{\rm w} / v_\infty$ & St. Dev. $\dot{M}_{\rm w} / v_\infty$ & Stellar type & Comparison types \\ \hline
IGR J18410-0535 & $0.122$ & $0.088$ & B1Ib/B0.2Ibp/B1I & B1Ia/B1Iab \\
SAX J1818.6-1703 & $0.124$ & $0.054$ & B0.5Iab & B0.5Ia \\
IGR J19140+0951 & $0.124$ & $0.054$ & B0.5Ia & B0.5Ia \\
IGR J18483-0311 & $0.123$ & $0.071$ & B0.5-B1 & B0.5Ia/B1Ia/B1Iab \\ \hline
\end{tabular}\\
\end{table*}

\subsection{Combining millimeter with bow shock constraints}

The SgXBs Vela X-1 and 4U 1907+09 are the two only known neutron star HMXBs that create an observed bow shock in the ISM\footnote{We note that the SFXT IGR J17544-2619 is runaway system as well \citep{maccarone2014}. Because it's proper motion is dominated by its radial motion, no bow shock is seen in IR, optical, or radio.}. A bow shock is formed when an object moves supersonically in the ISM, where the speed of sound for a standard density and temperature ($\sim 1$ cm$^{-3}$ and $\sim 10^4$ K) is of the order $10$ km/s. For runaway massive stars, ejected from their birth stellar population via dynamical or supernova-driven ejection mechanisms \citep{poveda1967,blaauw1961}, the bow shock is formed at a distance from the star: at this stand-off distance, the combined thermal and ram pressure of the ISM, moving at the stellar velocity in the frame of the star, equals the ram pressure of the wind moving at its terminal velocity. As a result, the stand-off distance $R_0$ can be related to the stellar wind parameters, peculiar stellar velocity $v_*$, and ISM properties (density $n_{\rm ISM}$, temperature $T_{\rm ISM}$), to write: 
\begin{equation}
\begin{split}
    \dot{M}_{\rm w} v_\infty &= 4\pi n_{\rm ISM}R_0^2 \left(m_p v_*^2 + kT_{\rm ISM}\right)\\
    &\approx 4\pi  m_p n_{\rm ISM}R_0^2 v_*^2,
\end{split}
\end{equation}
where $m_p$ is the proton mass and the approximation is accurate for $T_{\rm ISM} < 10^4$ K to within 2.7\% and 0.3\% for Vela X-1 \citep[$v_* \approx 55$ km/s;][]{gvaramadze2018} and 4U 1907+09 \citep[$v_* \approx 160$ km/s;][]{gvaramadze2011}, respectively. The stand-off distance is resolvable for both objects (0.57 pc and 0.89 pc, respectively), and hence the product of mass-loss rate and terminal velocity can be constrained.

\begin{figure}
\includegraphics[width=\columnwidth]{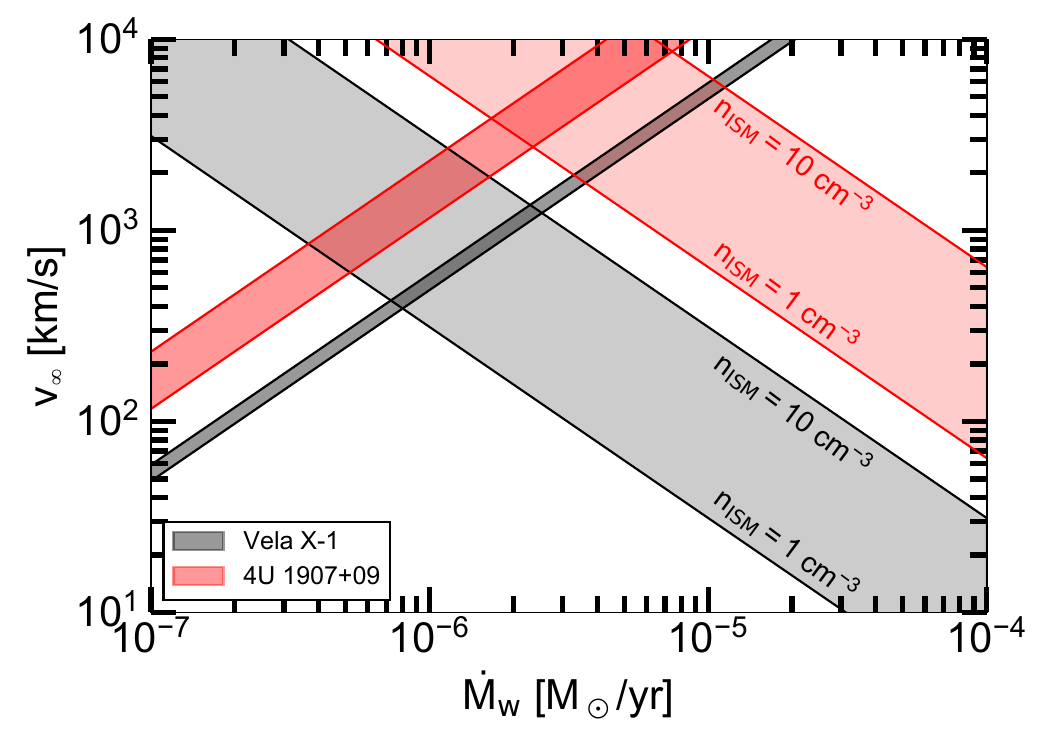}
 \caption{A comparison of the constraints on stellar wind mass loss rate and terminal velocity from millimeter observations (increasing curves) and bow shocks (decreasing curves). The latter is shown for a range of ISM densities, ignoring the second order effect of thermal ISM pressure.}
\label{fig:bowshocks}
\end{figure}

In Figure \ref{fig:bowshocks}, we show the regions in the mass-loss rate and terminal velocity space that are permitted for Vela X-1 and 4U 1907+09, based on their millimeter emission and their bow shocks. For the latter, we plot the constraints assuming an ISM density between 1 and 10 cm$^{-3}$. The band shown for the former represents the flux density uncertainty in the millimeter band and distance uncertainty. We find that the constraints for Vela X-1 intersect at $\dot{M}_{\rm w} \approx 8\times10^{-7}\text{ } (n_{\rm ISM} / \text{ 1 cm}^{-3})^{1/2}$ $M_\odot/$yr and $v_\infty = 4\times10^2\text{ } (n_{\rm ISM} / \text{ 1 cm}^{-3})^{1/2}$ km/s; for 4U 1907+09, we find $\dot{M}_{\rm w} \approx 2\times10^{-6} \text{ }(n_{\rm ISM} / \text{ 1 cm}^{-3})^{1/2}$ $M_\odot/$yr and $v_\infty = 3\times10^3\text{ } (n_{\rm ISM} / \text{ 1 cm}^{-3})^{1/2}$ km/s. These values for Vela X-1 are consistent with recent literature measurements (e.g., Table \ref{tab:appendix_literaturewinds}) for ISM densities of the order a few per cm$^3$ -- a density consistent with constraints from modelling its bow shock in the radio band \citep{vandeneijnden2022}. For 4U 1907+09, a density below 1 cm$^{-3}$ is needed to match earlier estimates in the literature. 

\subsection{Variability in low-frequency stellar wind emission}
\label{sec:var}

\begin{figure}
\includegraphics[width=\columnwidth]{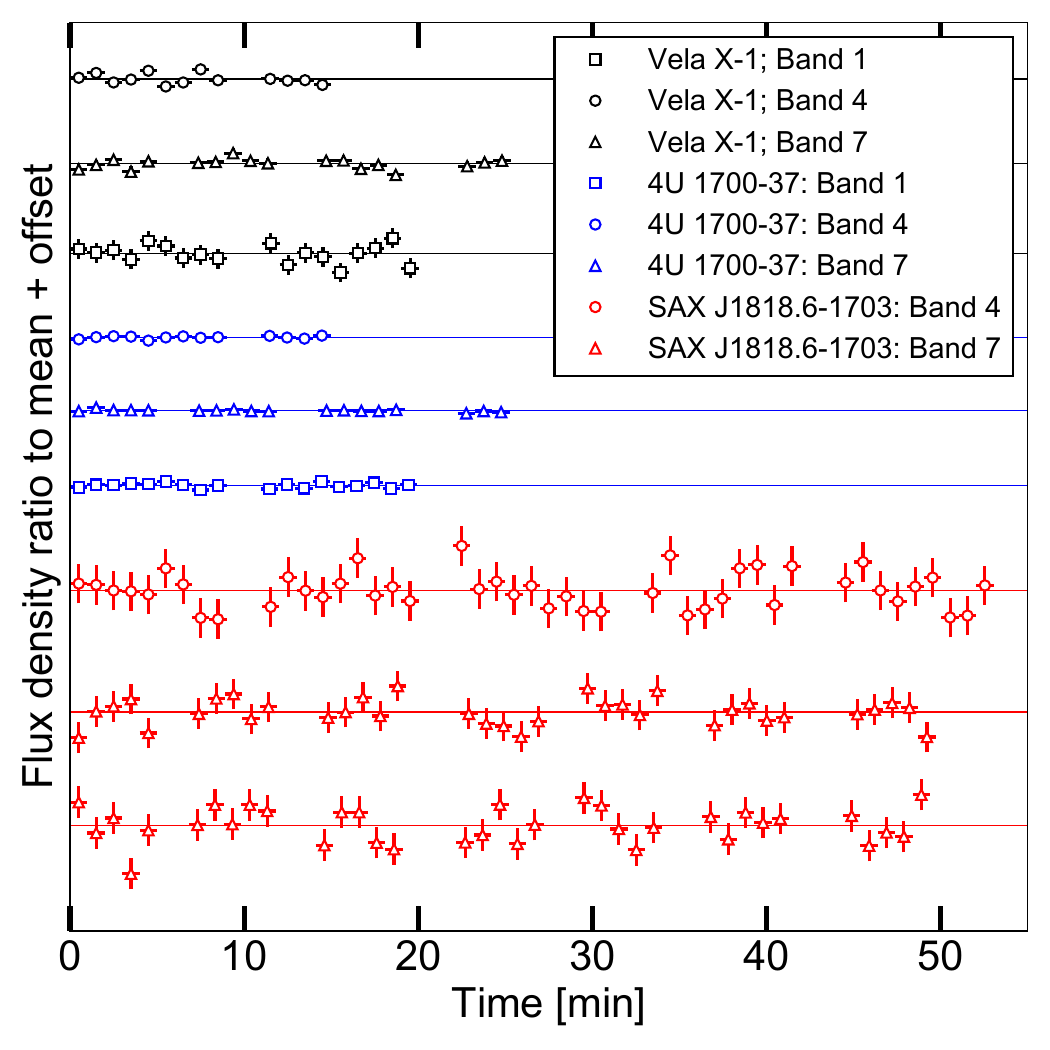}
\caption{The ALMA $1$-minute light curves for the targets and observing bands where the signal-to-noise is sufficient to detect the target in each time bin at $\geq 3\sigma$. Each light curve has been re-scaled by its mean and given an arbitrary vertical offset for clarity. Errors are shown at the $1\sigma$ level. The SAX J1818.6-1703 Band 7 light curve consists of two parts, as the full observation exceeded a single observing block.}
\label{fig:variability}
\end{figure}

A search for intra-observational variability in the ALMA observations did not reveal millimeter flaring nor flux density evolution on short ($\sim$ minutes) time scales. For a subset of targets and observing bands, that may not be surprising, given the insufficient signal-to-noise: we can estimate a minimum signal-to-noise in the time-integrated observation of length $T$ required for a $\gtrsim 3\sigma$ detection in each time bin of length $\Delta t$ and a constant source, as $3\sqrt{T/\Delta t}$. For, e.g., $\Delta t = 1$ minute, only 4U 1700-37, Vela X-1,  and Bands 4 and 7 of SAX J1818.6-0535 reach this requirement. For those sources/observing bands, we can calculate the reduced $\chi^2$ of a constant fit to the $1$-minute light curve as a first assessment of variability. In all cases, $\chi^2_{\nu,\rm constant} \leq 1$, implying the light curves are consistent with being constant. These light curves, re-scaled to their mean and shown with a vertical offset, are shown in Figure \ref{fig:variability}.

Alternatively, we can convert this non-detection of variability in an upper limit on fractional variability. The fractional variability $F_{\rm var}$, as defined by \citet{vaughan2003}, cannot be calculated for cases where $\chi^2_{\nu,\rm constant} \leq 1$, as this statement is equivalent to the variance $S^2$ being smaller than the mean square error $\overline{\sigma^2_{\rm err}}$ (see their Equation 9). To understand the minimum detectable fractional variability at $3\sigma$, we can assess when $F_{\rm var} > 3\times\text{err}(F_{\rm var})$. In the low-variability regime (e.g., the first case of Equation B1 in \citet{vaughan2003}), this is equivalent to a minimum detectable fractional variability of 
\begin{equation}
    F_{\rm var, detectable} \geq \left(\frac{9}{2N}\right)^{\frac{1}{4}} \frac{\sqrt{\overline{\sigma^2_{\rm err}}}}{\overline{S_{\nu}}}\text{ ,}
\end{equation}
\noindent where $N$ is the number of bins and $\overline{S_{\nu}}$ is the mean flux density in the light curve. Further assuming that the uncertainty in each time bin scales with $\sqrt{N}$, this limit is equivalent with 
\begin{equation}
    F_{\rm var,detectable} \geq \left(\frac{9}{2N}\right)^{\frac{1}{4}} \text{SNR}^{-1} \text{ ,}
\end{equation}
\noindent where SNR equals the signal-to-noise ratio of the full observation. This calculation implies the limits on the fractional variability at a $1$ minute time scale are 0.5\% / 0.4\% / 0.3\% for 4U 1700-37 (Band 1/4/7); 2\% / 1.5\% / 1\% for Vela X-1 (Band 1/4/7); and 2.1\% / 1.9\% for SAX J1818.6-0535 (Band 4/7). 

On time scales beyond individual observations, several approaches can be taken to assess the presence of variability. For instance, the comparison of the three ALMA bands, expected to lie on a single spectrum, did not reveal obvious variability between days of observations. On longer time scales, however, evidence for low-frequency variability can be found along different paths.

Firstly, IGR J18410-0535 was detected first with NOEMA in February 2023 at $63.4\pm9.6$ $\mu$Jy. The ALMA observations, all taking place in October 2024, revealed a spectrum at lower flux density; the SED fit returns a $100$ GHz range of $\sim 20$ -- $30$ $\mu$Jy. Given the systematic uncertainty on the NOEMA and ALMA flux density scale of the order $\lesssim 10$\%, such a factor $\sim 2$ variation appears intrinsic to the source. Secondly, in the radio band, 4U 1700-37 is observed at significantly lower flux densities ($\sim 250$ $\mu$Jy at 6 GHz) than during the 2018 ATCA observations ($484\pm13$ $\mu$Jy at 5.5 GHz) presented in \citet{vandeneijnden2021}. Again, a factor $\sim 2$ variability is seen, with a potentially shallower ($\alpha =  0.46 \pm 0.16$) radio spectrum in 2018 than in millimeter in 2024 ($\alpha =  0.70 \pm 0.01$ from ALMA and VLA). Finally, the Vela X-1 VLA radio spectrum appears potentially shallower ($\alpha = 0.36 \pm 0.33$) than the millimeter spectrum ($\alpha = 0.78 \pm 0.02$), with $\sim 12$ days separating the radio and millimeter observations. 

From the inverted spectra we earlier assumed that the same thermal wind emission underlies the low-frequency spectrum at all different epochs. For such a spectrum, changes in the emission can be driven by global changes in the wind mass loss rate or its velocity. To estimate the time scales of such variability, we can follow \citet{gudel2002} and write for the optically thick radius of the stellar wind

\begin{equation}
\begin{split}
    R_{\rm thick} = 3.6\times10^2 \text{} R_\odot & \left(\frac{\dot{M}_{\rm w}}{10^{-6} M_\odot/\text{yr}}\right)^{2/3} \left(\frac{v_\infty}{1000\text{ km/s}}\right)^{-2/3} \times \\
    & \left(\frac{T_e}{10^4\text{ K}}\right)^{-9/20} \left(\frac{\nu}{100\text{ GHz}}\right)^{-7/10}.
\end{split}
\end{equation}
Any significant change in the emission across such a scale, originating from a change in global parameters of the wind, will require at least a causal time scale of variability $\tau_{\rm thermal} \propto R_{\rm thick} / v_\infty$, which we can write as

\begin{equation}
\begin{split}
\tau_{\rm thermal} \approx 2.9 \text{ days} &\left(\frac{\dot{M}_{\rm w}}{10^{-6} M_\odot/\text{yr}}\right)^{2/3} \left(\frac{v_\infty}{1000\text{ km/s}}\right)^{-5/3} \times \\
&\left(\frac{T_e}{10^4\text{ K}}\right)^{-9/20} \left(\frac{\nu}{100\text{ GHz}}\right)^{-7/10}.
\end{split}
\end{equation}
Variability on days time scale, and beyond, in the millimeter band can therefore be consistent with changes in the large-scale thermal emission reflecting changes in global wind parameters. 

The stellar winds of B-type supergiants are known to be variable from studies focusing on singles or non-degenerate binaries. Such studies often focus on the detection of variability periods in photometry or, alternatively, H$\alpha$ properties \citep{kaufer2006,lefever2007,lefevre2009}, which have been linked to opacity-driven oscillations in the star \citep{kraus2015}. Radial pulsations have, in particular, been linked to variability in the mass loss rates of the star \citep{glatzel1999,yadav2016}. In addition, as the stellar winds are radiatively driven, instabilities in the wind driving mechanism may further lead to variable outflow rates. Focusing on measurements of the wind parameters instead of variability, significant changes have indeed been observed: while \citet{prinja1986} reported variations in mass-loss rates of $\sim 10$\% on $\geq$days time scales, \citet{lefever2007}, \citet{kraus2015}, and \citet{haucke2018} report changes by factors $1.05$ to $2.7$ between observations separated by a day to $\sim$a year. Alternatively, \citet{kaper1996} and \citet{massa2024} performed systematic UV studies of a sample of ten and twenty-five OB stars, respectively, with \citet{massa2024} for instance reporting an average $1\sigma$ variation in mass-loss rate of $22$\% -- although individual stars range between $8$\% and $45$\%. 

The analyses above focus on the wind variability from diagnostics probing into the inner wind regions. However, if such changes propagate to the large-scale millimeter and radio emitting regions, their magnitude would be sufficient to explain factor $\sim 2$ changes in the low-frequency SED normalization in SgXBs. This approach ignores the complex environment created by the presence of the compact object; the presence of large-scale wakes, and potential induced variability in the wind structure. The small number of measurements of the low-frequency SgXB emission in this work, prevents a more detailed analysis taking into account such effects; in a future work, we aim to analyse these processes in the radio/millimeter variability of SgXBs in more detail. 

\section{Conclusions}
\label{sec:conclusions}

We performed a first systematic millimeter study of a sample of neutron star HMXBs, focusing on twelve SgXBs and SFXTs using ALMA, NOEMA, and the VLA. In this work, we presented the results from these observations and focused in particular on the origin of their low-frequency emission and on the comparison between the two subclasses of targets. From our work, we draw the following conclusions:

\begin{enumerate}
\item Using a combination of ALMA, NOEMA, and VLA observations, nine out of the twelve SgXBs and SFXTs were detected in at least one observing band. Two of the non-detected targets are SgXBs, while the third is an intermediate source. These non-detections are not explained merely by distance (Table \ref{tab:mmresults})
\item Based on the spectral shapes and specific luminosities of the targets, we conclude that the radio and millimeter emission of neutron star SgXB and SFXTs is dominated by thermal free-free stellar wind emission. 
\item In neutron star SgXBs and SFXTs, jet emission is unlikely to be detected, unless an SED is measured with a sufficient number of bands down to sub-GHz frequencies so that a flattening of the spectrum towards low frequencies can be measured. However, even then a favorable binary configuration and/or brightening of the jet compared to Be/X-ray binaries due interactions with wind material, is required to observe these jet signatures.
\item The millimeter band shows evidence for systematic differences between the stellar winds in SFXTs and SgXBs. In particular, the four SFXTs discussed in this work are intrinsically fainter at 100 GHz than prototypical SgXBs (e.g., Vela X-1, 4U 1700-37, 2S 0114+650) and SgXB detected in the radio band priorly (GX 301-2, IGR J16318-4848, and IGR J16320-4751). As these SgXBs also show typically stronger Fe K$\alpha$ lines, we find an agreement between probes of the stellar wind density on global scales and local scales close to the neutron star. A \textit{globally} denser (e.g., because slower) stellar wind in SgXBs could explain both effect qualitatively, which we will explore quantitatively in a companion work.
\item The observed low-frequency spectra are slightly more inverted than the analytical stellar wind spectral shapes predicted by \citet{wright1975} and \citet{panagia1975}. Applying the analytical prescriptions nonetheless, we find constraints on mass-loss rate over wind velocity that are inconsistent with the literature for several sources. Wind clumping, wind variability, and progressive wind acceleration may underlie these differences.
\item For Vela X-1, the millimeter constraints on the wind parameters agree with geometrical bow shock constraints for reasonable ISM densities. For 4U 1907+09, low ISM densities ($n_H < 1$ cm$^{-3}$) are required.
\item No fast millimeter variability (e.g., within observations) is seen down to low values of the fractional variability, as expected for thermal wind emission. On longer time scales (e.g., at least several days), where the wind may causally variable, we find evidence for variability that warrants dedicated follow-up observations for further confirmation.
\end{enumerate}

\section{Acknowledgments}
We thank the referee for their constructive review of this manuscript. For the purpose of open access, the authors have applied a Creative Commons Attribution (CC-BY) license to any Author Accepted Manuscript version arising from this submission. The authors thank all telescope operators for their work during the observations underlying this research; Edwige Chapillon for their assistance in NOEMA data reduction and imaging; and Karran Kumar for discussions about runaway high-mass X-ray binaries in Gaia DR3. JvdE is supported by funding from the European Union's Horizon Europe research and innovation programme under the Marie Skłodowska-Curie grant agreement No 101148693 (MeerSHOCKS) and acknowledges a Warwick Astrophysics prize post-doctoral fellowship made possible thanks to a generous philanthropic donation. This paper makes use of the following ALMA data: ADS/JAO.ALMA\#2024.1.00657.S. ALMA is a partnership of ESO (representing its member states), NSF (USA) and NINS (Japan), together with NRC (Canada), NSTC and ASIAA (Taiwan), and KASI (Republic of Korea), in cooperation with the Republic of Chile. The Joint ALMA Observatory is operated by ESO, AUI/NRAO and NAOJ. The National Radio Astronomy Observatory is a facility of the National Science Foundation operated under cooperative agreement by Associated Universities, Inc. This work is based on observations carried out under project number w23bo with the IRAM NOEMA Interferometer. IRAM is supported by INSU/CNRS (France), MPG (Germany) and IGN (Spain). The research leading to these results has received funding from the European Union’s Horizon 2020 research and innovation program under grant agreement No 101004719 [Opticon RadioNet Pilot ORP]. This research has made use of NASA's Astrophysics Data System Bibliographic Services and the SIMBAD database, operated at CDS, Strasbourg, France. This work made use of \textsc{Astropy}\footnote{http://www.astropy.org}, a community-developed core Python package and an ecosystem of tools and resources for astronomy \citep{astropy:2013, astropy:2018, astropy:2022}, \textsc{emcee} \citep{emcee2013}, and \textsc{numpy} \citep{harris2020array}.

\section*{Data Availability}
A \textsc{GitHub} reproduction repository will be made public upon publication at \url{https://github.com/jvandeneijnden/mmwinds1}, containing all files to repeat the data reduction and analysis presented in this paper. A reproduction package will also be released via \textsc{zenodo} via the following DOI: \url{10.5281/zenodo.17357078}. Raw data associated with this work will be available in their respective observatory archives after a one-year proprietary period. Earlier access after acceptance of the paper may be granted by the PI upon reasonable request. 

\bibliographystyle{mnras}
\bibliography{main.bib}

\appendix

\section{MM images}

\begin{figure*}
\includegraphics[width=\textwidth]{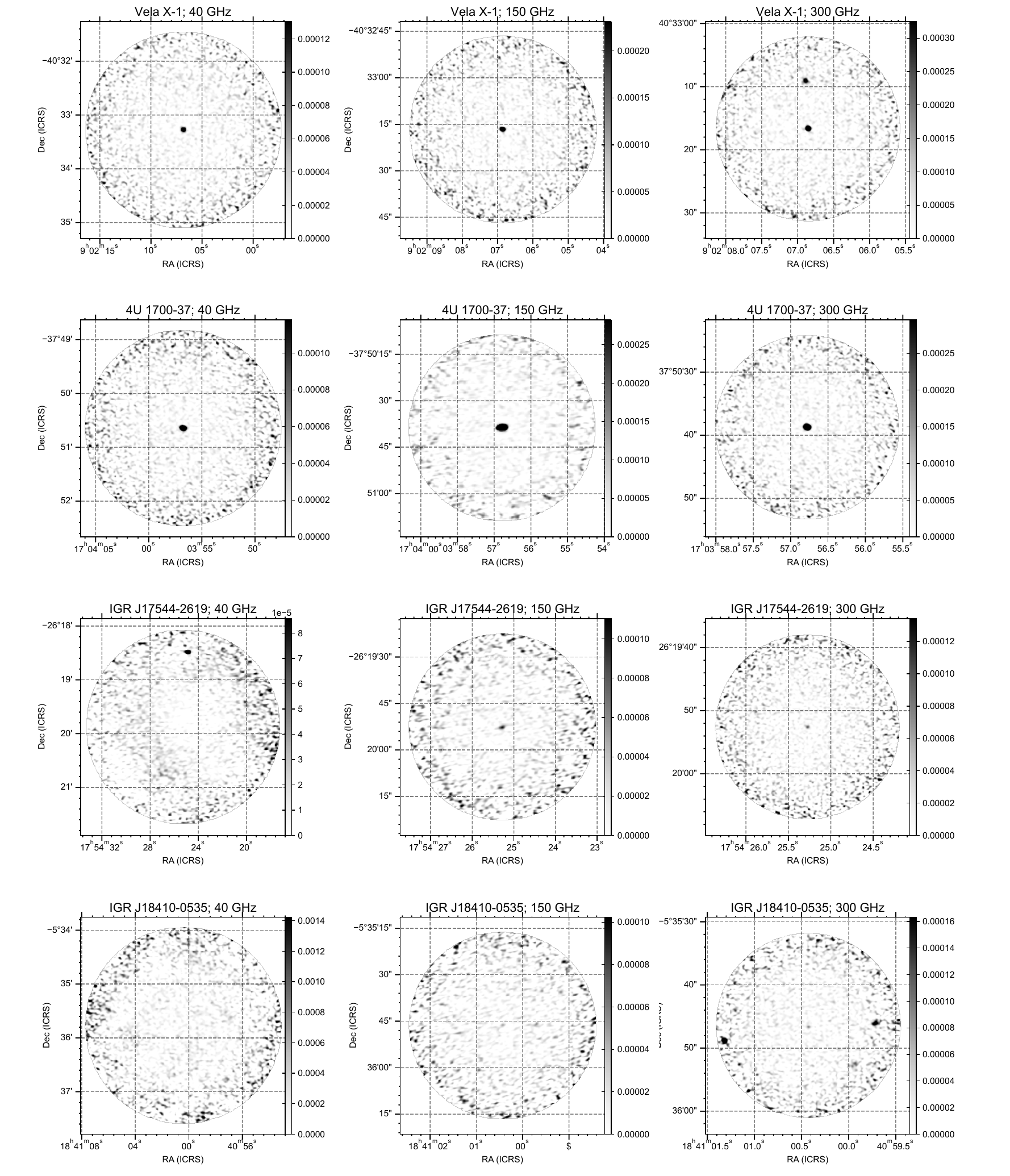}
 \caption{The individual reduced millimeter images of all targets (except 4U 1907+09 at 40/100/150 GHz, shown in the main paper), in each observing band. The target and representative frequency are shown in each subplots title. Continued in Figure \ref{fig:all2}.}
\label{fig:all1}
\end{figure*}

\begin{figure*}
\includegraphics[width=\textwidth]{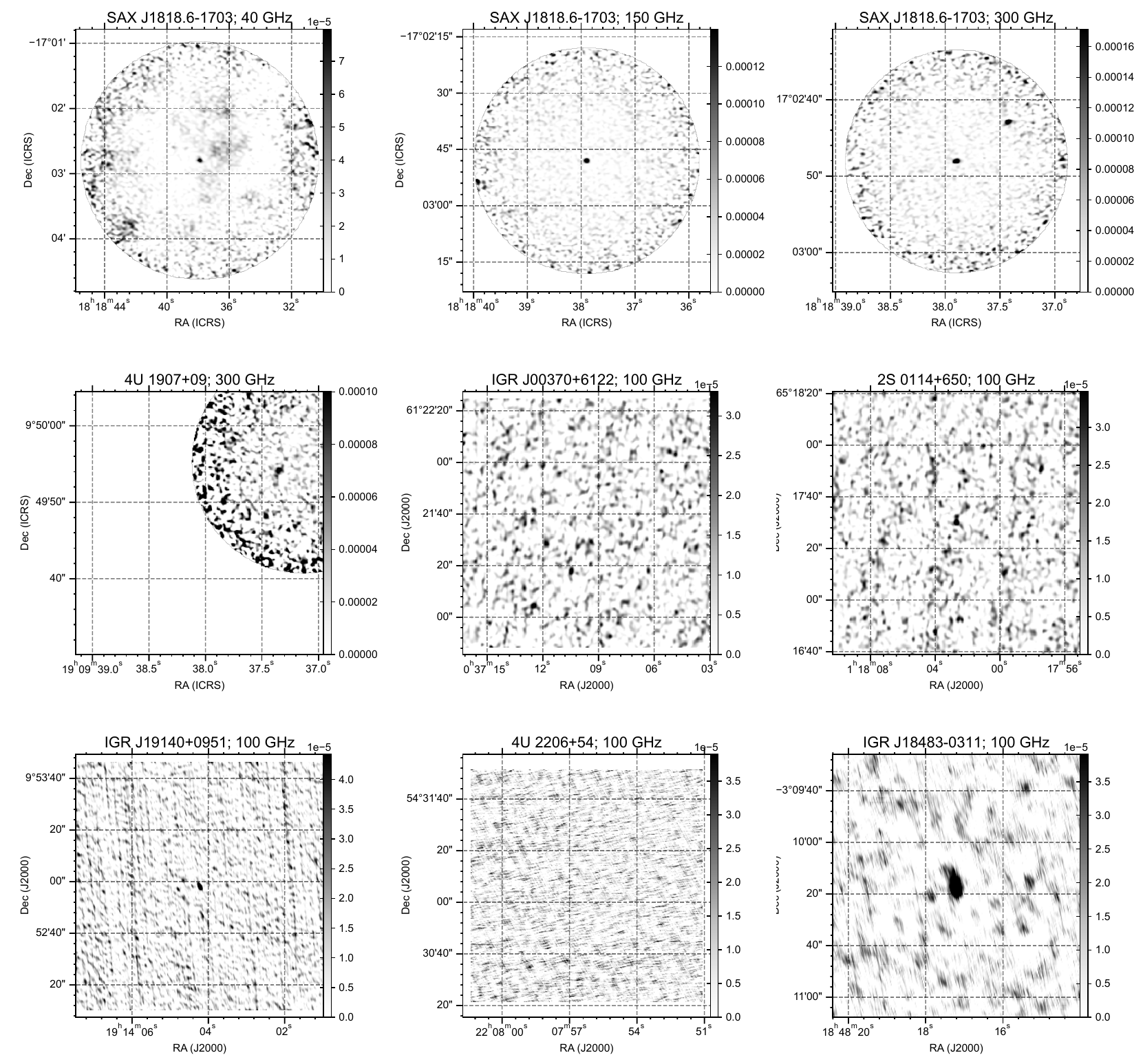}
 \caption{Same as Figure \ref{fig:all1}.}
\label{fig:all2}
\end{figure*}

%%%%%%%%%%%%%%%%%%%%%%%%%%%%%%%%%%%%%%%%%%%%%%%%%%

% Don't change these lines
\bsp	% typesetting comment
\label{lastpage}
\end{document}